\documentclass[]{aa}  

\usepackage{graphicx}
\usepackage{txfonts}
\begin{document}

   \title{High-resolution ALMA observations of transition disk candidates in Lupus}
\titlerunning{High-res observations of transition disk candidates}

\author{Nienke van der Marel
\inst{1,2}
\and
Jonathan P. Williams
\inst{3}
\and
Giovanni Picogna
\inst{4}
\and
Sierk van Terwisga
\inst{5}
\and
Stefano Facchini
\inst{6}
\and
Carlo F. Manara
\inst{7}
\and
Apostolos Zormpas
\inst{4}
\and
Megan Ansdell
\inst{8}
}

\institute{Leiden Observatory,
Leiden University,
P.O. Box 9531, 
NL-2300 RA Leiden, 
the Netherlands
\and Physics \& Astronomy Department, 
University of Victoria, 
3800 Finnerty Road, 
Victoria, BC, V8P 5C2, 
Canada
\and Institute for Astronomy, University of Hawai'i at Manoa, Woodlawn Dr., Honolulu, HI 2680, USA
\and University Observatory, Faculty of Physics, Ludwig-Maximilians-Universität München, Scheinerstr. 1, 81679, Munich, Germany
\and Max-Planck-Institut für Astronomie, Königstuhl 17, 69117, Heidelberg, Germany
\and Dipartimento di Fisica, Università degli Studi di Milano, Via Celoria 16, I-20133 Milano, Italy
\and European Southern Observatory, Karl-Schwarzschild-Str. 2, D-85748 Garching, Germany
\and NASA Headquarters, 300 E Street SW, Washington, DC 20546, USA
}

   \date{Submitted on March 1st, 2022}

 
   \abstract
   {Transition disks with small inner dust cavities are interesting targets for the study of disk clearing mechanisms. Such disks have been identified through a deficit in the infrared part of their SED, but spatially resolved millimeter imaging is required to confirm the presence of an inner dust cavity.}
   {The aim of this study is to image the millimeter-dust distribution of 10 transition disk candidates selected from both SED criteria and from tentative detections of inner cavities in low-resolution observations to verify their nature and to understand their origin.}
   {
We use high-resolution ALMA observations of 30 mas resolution in Band 6 continuum and $^{12}$CO 2--1 emission of 10 transition disk candidates in the Lupus star forming region, in order to confirm the presence of inner dust cavities and infer the responsible mechanism. The continuum data are analyzed using visibility modeling and the SEDs are compared with radiative transfer models.}
   {Out of the six transition disk candidates selected from their SED, only one disk revealed an inner dust cavity of 4 au in radius. Three of the other disks are highly inclined, which limits the detectability of an inner dust cavity but it is also demonstrated to be the possible cause for the infrared deficit in their SED. The two remaining SED-selected disks are very compact, with dust radii of only $\sim$3 au. From the four candidates selected from low-resolution images, three new transition disks with large inner cavities $>$20 au are identified, bringing the total number of transition disks with large cavities in Lupus to 13.}
   {SED-selected transition disks with small cavities are biased towards highly inclined and compact disks, which casts doubt on the use of their occurence rates in estimating dispersal timescales of photoevaporation. Using newly derived disk dust masses and radii, we re-evaluate the size-luminosity and $M_{\rm dust}-M_{\rm star}$ relations. These relations can be understood if the bright disks are dominated by disks with substructure whereas faint disks are dominated by drift-dominated disks.}


   \maketitle

\section{Introduction} 
\label{sec:intro}
One of the key questions in planet formation studies is the dispersal of protoplanetary disks. Transition disks with inner dust cavities have long been hauled as the intermediate regime between optically thick protoplanetary and optically thin debris disks, 'caught in the act' of disk dispersal \citep{WilliamsCieza2011}. Transition disks were originally identified through a dip in the infrared part of their SED \citep{Strom1989,Calvet2004}, where the largest inner cavities of $>$20 au radius were confirmed through millimeter interferometry \citep[e.g.][]{Brown2009,Andrews2011,Francis2020}. However, the high stellar accretion rates in combination with large cavity sizes measured in such systems are incompatible with models of photoevaporative clearing \citep[e.g.][]{OwenClarke2012}, which is thought to be the main dispersal mechanism of protoplanetary disks \citep{Clarke2001,Ercolano2017}. It has been suggested that two types of transition disks exist \citep{OwenClarke2012}, caused by either photoevaporation or by other mechanisms such as clearing by a wide-orbit companion \citep{LinPapaloizou1979}, where the latter is the most likely explanation for the large cavity transition disks with high accretion rates. More advanced photoevaporation models including carbon depletion \citep{Ercolano2018,Wolfer2019}, updated models of X-ray radiation \citep{Picogna2019} and dead zones \citep{Garate2021} can reproduce a larger range of observed transition disks, including the ones with high accretion rates and large cavities, but not yet at the observed fraction.  However, these comparisons still mostly rely on transition disk candidates with small $<$10 au cavities identified through SED modeling \citep[e.g.][]{Merin2010}, while these cavities have so far not been confirmed with millimeter imaging. 

Large cavity transition disks have received much attention in both ALMA and scattered light imaging studies \citep[for an overview see e.g.][]{Francis2020,Andrews2020}. Although wide planetary companions have so far only been confirmed in one transition disk to date \citep{Keppler2018,Haffert2019}, indirect evidence for companions at $\sim$10-20 au orbits in transition disks has been inferred from e.g. deep gas gaps \citep{vanderMarel2016-isot,Dong2017}, spiral density waves \citep{Muto2012,Dong2015spirals} and misalignments between inner and outer disk \citep{Marino2015,Benisty2017,Bohn2021}. The constraints on detectable companions through high contrast imaging are still limited to several Jupiter masses in most disk cavities, in particular closer to the host star \citep{vanderMarel2021asym,Asensio2021}. For older systems, Jovian companions at wide orbits $>$20 au appear to be rare \citep{Nielsen2019,Vigan2021}, but this could be mitigated by considering the stellar mass dependence and inward migration of such planets  \citep{vanderMarelMulders2021}, as giant planets are primarily located at 3-8 au \citep{Fernandes2019,Fulton2021}.

On the other hand, transition disks with small inner dust cavities with cavity radii $<$10 au have received much less attention in ALMA studies, as these disks tend to be more compact and fainter and resolving their inner cavities requires very long baseline observations. Such disks are interesting to study to test photoevaporation models, but also for possible small scale cavities carved by giant planets at small orbits of a few au which have formed in situ. The only small cavity transition disk candidates studied at high resolution to date are CX Tau \citep{Facchini2019} which did not reveal an inner cavity, the ISO-Oph~2 secondary with a 2.2 au inner cavity \citep{Gonzalez2020} and XZ Tau B \citep{Osorio2016} which revealed an inner cavity of 1.3 au radius, but this was not recovered in later high resolution observations \citep{Ichikawa2021}. Small inner cavities $\lesssim$10 au were resolved in millimeter images of Sz~129 and MHO~6, despite not showing any indication of a dip in the SED  \citep{Andrews2018,Kurtovic2021}. A systematic study of small cavity transition disk candidates and their occurrence is still lacking.

The young Lupus association ($\sim$160 pc, $\sim$2 Myr old) is one of the best-studied nearby star forming regions. In particular its protoplanetary disk population has been studied in complete multi-wavelength ALMA continuum surveys \citep{Ansdell2016,Ansdell2018,Tazzari2017,Tazzari2020,Tazzari2021,Sanchis2020}, stellar properties \citep{Alcala2014,Alcala2017,Alcala2019}, transition disks with large cavities \citep{vanderMarel2018a,Tsukagoshi2019-sz91}, gas disk masses and sizes \citep{Miotello2017,Miotello2021,Trapman2020,Sanchis2021} and chemistry studies \citep{vanTerwisga2019,Miotello2019}. In this work we will use the stellar properties and the continuum properties in connection with predictions from dust evolution models to place the transition disk candidates into context. A number of transition disk candidates have been identified in Lupus \citep{Merin2010,Bustamante2015,vanderMarel2016-spitzer} which have only partially been confirmed with millimeter imaging for the largest cavities \citep{vanderMarel2018a}. This region is thus well suited for a detailed study of transition disk candidates with small cavities. 

In this work we present a high-resolution sample study of transition disk candidates in the Lupus star forming region, imaged with ALMA at an angular resolution of $\sim$30 mas ($\sim$5 au). In Section \ref{sec:sample} we present the sample selection and the technical details of the observations. Section \ref{sec:results} presents the initial images and Section \ref{sec:model} the visibility modeling procedure and best-fit results used to describe the intensity profiles of the disks, together with the radiative transfer modeling procedure used to match the SEDs. Section \ref{sec:discussion} discusses the results in context of target selection and possible consequences for disk and dust evolution, which are summarized in Section \ref{sec:conclusions}.

\section{Observations}
\label{sec:obs}

\subsection{Sample}
\label{sec:sample}
A full sample of transition disk candidates in Lupus was constructed in \citet[][Table 2]{vanderMarel2018a}. Out of the 95 Lupus disk members identified by \citet{Ansdell2016,Ansdell2018}, 11 disks showed evidence for an inner dust cavity in the ALMA Band 7 continuum observations at $\sim$0.25" resolution, the spatial resolution of the initial Lupus disk survey. These 11 disks were analyzed and their cavity sizes were presented in \citet{vanderMarel2018a}, with cavity sizes ranging from 15--90 au. However, for 6 transition disk candidates from previous SED studies \citep{Bustamante2015,vanderMarel2016-spitzer} the inner cavity remained unresolved in the Lupus disk survey. These candidates were identified based on SED analysis of \emph{Spitzer} data of YSOs in complete Lupus disk surveys. Furthermore, 2 transition disk candidates were suggested from visibility modeling of the Lupus Band 7 data by \citet{Tazzari2017}, which suggest a depletion in the inner part of the disk, although not resolved in the image plane. These 8 targets thus are the most promising candidates for studying disks with small cavities $<$20 au in radius in the Lupus star forming region. 

In addition, 2 targets were identified as potential edge-on transition disks with large cavities $>$25 au by \citet{Ansdell2018}, although the low spatial resolution data could not confirm the morphology: J16090141 and J16070384. These disks were included in the observing sample as well, although they are clearly of a different nature than the other 8 transition disk candidates.

The full sample is presented in Table \ref{tbl:sample}, together with its stellar properties. Stellar properties were initially derived using X-shooter spectroscopy by \citet{Alcala2017}, and updated for the \emph{Gaia} DR2 distances by \citet{Alcala2019}.

\begin{table*}[!ht]
\footnotesize
\begin{center}
\caption{Sample}
\label{tbl:sample}
\begin{tabular}{llllllrlll}
\hline
Target&Coordinates&SpT$^a$&$L_*^a$&$M_*^a$&$A_V^a$&$\log M_{\rm acc}^a$&$d^b$&Type&Ref$^c$\\
&&&($L_{\odot}$)&($M_{\odot}$)&&($M_{\odot}$ yr$^{-1}$)&(pc)&\\
\hline
Sz76	&	15:49:30.716 -35:49:51.890	&	M4	&	0.18	&	0.23	&	0.2	&	-9.3	&	160	&		Small cavity SED	&	1	\\
Sz103	&	16:08:30.254 -39:06:11.647	&	M4	&	0.22	&	0.23	&	0.7	&	-9.0	&	160	&		Small cavity SED	&	1	\\
Sz104	&	16:08:30.799 -39:05:49.320	&	M5	&	0.07	&	0.23	&	0.0	&	-9.8	&	165	&		Small cavity SED	&	1	\\
Sz112	&	16:08:55.512 -39:02:34.420	&	M5	&	0.12	&	0.17	&	1.0	&	-9.7	&	160	&		Small cavity SED	&	1	\\
J16011549	&	16:01:15.529 -41:52:35.615	&	-$^d$	&		-&	-&	 -	&	$-$	&	160	&		Small cavity SED	&	1	\\
J16081497	&	16:08:14.959 -38:57:14.975	&	M5.5	&	0.01	&	0.10	&	0.0	&	-10.3	&	159	&		Small cavity SED	&	2	\\
J16000236	&	16:00:02.339 -42:22:15.055	&	M4	&	0.18	&	0.14	&	1.4	&	-9.7	&	164	&		Small cavity visibility	&	3	\\
Sz129	&	15:59:16.454 -41:57:10.759	&	K7	&	0.60	&	0.78	&	0.9	&	-8.4	&	162	&		Small cavity visibility	&	3	\\
J16090141	&	16:09:01.399 -39:25:12.398	&	M4	&	0.15	&	0.23	&	0.5	&	-9.6	&	164	&		Edge-on large cavity	&	4	\\
J16070384	&	16:07:03.829 -39:11:11.900	&	M4.5	&	0.02	&	0.16	&	0.6	&	-12.5$^e$&	159	&		Edge-on large cavity	&	4	\\

\hline
\end{tabular}
\end{center}
$^a$ All stellar properties are derived in \citet{Alcala2014}, and corrected for Gaia distances in \citet{Alcala2019}. \\
$^b$ Distances are taken inverting parallaxes of Gaia DR2.\\
$^c$ Reference for the identification of the transition disk candidate: 1) \citet{vanderMarel2016-spitzer}; 2) \citet{Bustamante2015}; 3) \citet{Tazzari2018}; 4) \citet{Ansdell2018}.\\
$^d$ Stellar properties unknown: SED consistent with M5, $L_*$=0.05$L_{\odot}$, $A_V$=17 and $M_*$=0.1$M_{\odot}$.\\
$^e$ Reported as sub-luminous by \citet{Alcala2017}, so accretion rate less reliable.	
\end{table*}

\subsection{ALMA observations}
\label{sec:almaobs}
The disks in the sample were observed in summer 2019 in ALMA Cycle 6 program 2018.1.01054.S (PI van der Marel) in Band 6. The 3 disks that were known to be extended in dust \citep[$>$0.5" radius,][]{Ansdell2018}, namely Sz~129, J16000236 and J16011549, were observed in a dual configuration setting with short baseline observations ($\sim$40-2,000m) on August 31th 2019 and long baseline observations ($\sim$200-13,000m) on June 6th 2019. The other 7 disks were already constrained to be $<$0.3" in size and could be observed in a single long baseline configuration ($\sim$200-13,000m) without losing flux. These disks were observed in multiple execution blocks on June 6, 18 and 19th 2019, with the exception of Sz76 which was observed in a single execution block on June 12th 2019. As the flux of each disk varied from 1-50 mJy per source, the sensitivity and integration time were set accordingly for each source. Integration times ranged between 8 and 20 minutes per source. The final rms noise levels are given in Table \ref{tbl:obs}.

The spectral settings were set to cover both the 230 GHz continuum and the CO 2--1 isotopologues at low spectral resolution to increase the continuum sensitivity, centered on 230.5, 231.5, 219.5 and 217.5 GHz, respectively. The 230.5 and 219.5 GHz spectral windows were set at 0.7 and 1.5 km s$^{-1}$ spectral resolution (564 and 1129 kHz), respectively, the other windows were set at TDM mode, resulting in a total continuum bandwidth of 6.1 GHz.

The data were calibrated using the provided pipeline reduction scripts using CASA 5.4. J1517-2422 or J1427-4206 were used as flux and bandpass calibrator and J1610-3958 or J1534-3526 as phase calibrator on the various executions. 

The calibrated datasets were concatenated and averaged over all channels to obtain continuum measurement sets. For Sz129, the short baseline data were recentered using \texttt{fixplanets} before concatenation with the long baseline data, the other targets were already positioned consistently with the long baseline data. Continuum images were obtained using the \texttt{tclean} CASA task with the \texttt{multiscale} deconvolver. The visibilities were weighed using briggs weighting with robust=0.5 when possible: for images with low image quality natural weighting was used and for J16070384 tapering was used to recover the image structure as the disk remained undetected in the natural resolution. Typical beam sizes of the final images are $\sim$0.03" or 5 au diameter at the distance of Lupus. The total flux in the image is measured and compared with the previously measured fluxes at low resolution by \citet{Ansdell2018}. Self-calibration was attempted for the brightest sources but did not improve the SNR. The imaging parameters and resulting beam size and noise levels are given in Table \ref{tbl:obs}. The final images are shown in Figure \ref{fig:gallery}.

\begin{table*}[!ht]
\caption{Properties ALMA images}
\label{tbl:obs}
\begin{tabular}{llllll|lll}
\hline
\multicolumn{6}{c|}{Continuum}&\multicolumn{3}{c}{$^{12}$CO line cubes}\\
\hline
Target&Weighting&Beam size&rms&peak SNR&$F_{1.3mm}^{\rm total}$  & Beam size & rms & peak SNR\\
&&(")&(mJy beam$^{-1}$)&&(mJy)&(")&(mJy km s$^{-1}$)&\\
\hline
Sz76 	& Briggs (r=0.5)	& 0.026$\times$0.025	& 0.019	&14	&3.4 & 0.19$\times$0.17 & 6.3 & 6 \\ 
Sz103	& Briggs (r=0.5)	& 0.026$\times$0.026	& 0.026	&25	&4.5 & -&-&- \\  
Sz104	& Briggs (r=0.5)	& 0.027$\times$0.026	& 0.026	&24	&1.1 & -&-&- \\  
Sz112	& Briggs (r=0.5)	& 0.027$\times$0.026	& 0.027	&16	&1.5 & -&-&- \\ 	
J16011549	& Natural 			& 0.040$\times$0.031	& 0.029	&75	&33 & 0.26$\times$0.23 & 5.6 & 21\\ 
J16081497& Natural			& 0.038$\times$0.033	& 0.025	&17	&3.4 & -&-&- \\ 
J16000236& Natural 			& 0.040$\times$0.031	& 0.028	&29	&58 & 0.26$\times$0.23 & 5.0 & 14 \\
Sz129	& Natural 			& 0.040$\times$0.031	& 0.032	&27	&74 & 0.26$\times$0.23 & 4.8 & 13 \\
J16090141	& Natural			& 0.038$\times$0.033	& 0.024	&11	&5.3 & 0.38$\times$0.36 & 12 & 10 \\ 
J16070384& Tapering (0.1")		& 0.125$\times$0.116		& 0.040	&5	&1.1 & -&-&- \\ 
\hline
\end{tabular}
\end{table*}

In addition, we attempted to image the $^{12}$CO 2--1 line cubes when detectable, after subtracting the continuum in the visibilities using the \textit{uvcontsub} task. The $^{12}$CO line could only be detected for half of the sources by intensive tapering to 0.2-0.3" resolution and a spectral resolution of 1 km s$^{-1}$. Spatial filtering likely affected the $^{12}$CO emission of the disks that were only observed at long baselines. Other lines than $^{12}$CO were not detected. The properties of the resulting image cubes are given in Table \ref{tbl:obs} and the moment maps are presented in Figure \ref{fig:COmaps}.

\section{Results}
\label{sec:results}

\subsection{Dust continuum}
\label{sec:continuum}
\begin{figure*}[!ht]
\centering
\includegraphics[width=1.1\textwidth,trim=40 0 20 0]{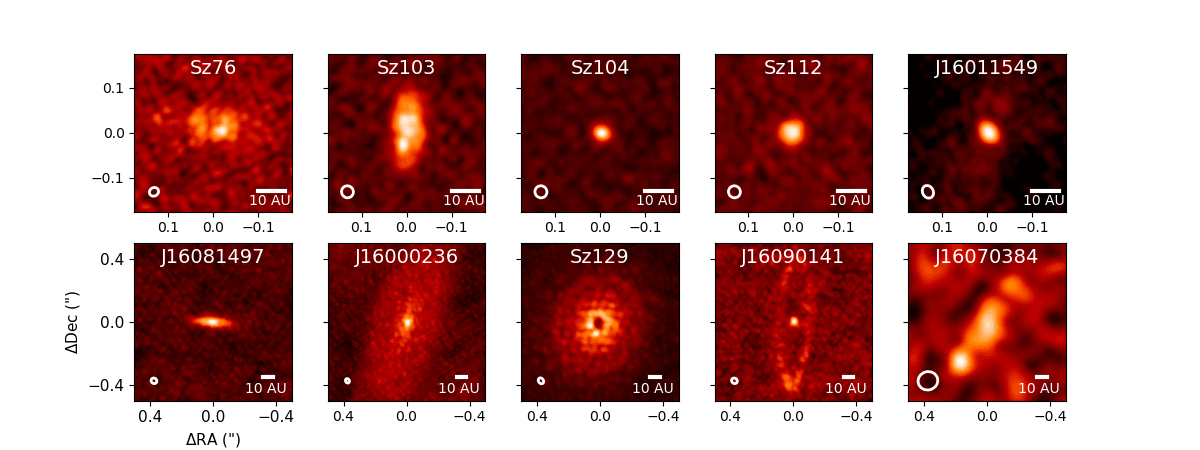}
\caption{Gallery of Band 6 continuum images of the entire sample. The beam size is indicates in the lower left corner. The top row is zoomed in further than the bottom row as the disks are smaller in size. The image properties are given in Table \ref{tbl:obs}. J16011549 is shown with an arcsin stretch to reveal the fainter outer disk. The first six disks were selected based on their SED, the other four based on a hint of an inner cavity in low-resolution observations.}
\label{fig:gallery}
\end{figure*}

Figure \ref{fig:gallery} presents the continuum images of the disks in our sample. All disks except Sz104 are spatially resolved, and Sz112 is marginally resolved. The majority of the disks appears to be highly inclined, which suggests at least for the first 6 disks that SED-derived transition disk candidates could be the result of inclination. All images recover the previously estimated 230 GHz continuum flux from \citet{Ansdell2018} within 5\%, with the exception of Sz76 and J16011549. The image of Sz76 recovers only 75\% of the flux, whereas the image of J16011549 overestimates the flux by 50\% compared to the previous measurements. The former may be the result of lack of short baselines, the latter is likely caused by the morphology of this disk, which consists of a very bright inner part and a much fainter extended disk: the surface brightness of the extended structure may have been below the detection limit of the low resolution data. This disk was observed by a combination of short and long baselines and is thus unlikely affected by lack of baseline coverage. 

The continuum images reveal a variety of structures based on visual inspection. Sz129 is the only disk with a clear inner dust cavity of 0.05" (8 au), directly revealed by its face-on nature, confirming its initial discovery in DSHARP by \citet{Huang2018}. The SNR is somewhat lower than the DSHARP data, but the total flux is consistent (within a flux calibration uncertainty of 10\%). Sz76 appears to show an inner depletion with a dust ring of $\sim$0.03" (5 au) radius, whereas Sz103 might have a dust ring with a central blob as well, although the image fidelity is fairly low. Sz104 and Sz112 are very compact with outer radii of $\lesssim$0.03" and do not show any evidence for an inner gap. J16011549 and J16000236 show a 'fried-egg' structure with a bright central part and extended emission similar to Sz98 and V1094~Sco \citep{vanTerwisga2018}, at moderately high inclination, and J16081497 is highly inclined without observable evidence for an inner depletion. The two disks marked as large cavity edge-on disks by \citet{Ansdell2018} indeed reveal large dust rings. The J16090141 disk shows a large dust ring with a radius of 0.4" (66 au), with a remarkably bright central feature, most likely the inner dust disk, which has a striking resemblance to the morphology of the Formalhaut debris disk \citep{MacGregor2017}. After correcting for the much closer distance of 7.7 pc of Formalhaut, J16090141 is 100 times more luminous in dust emission, consistent with its less evolved stage. The nature of the other highly inclined disk, J16070384, remains debatable due to the low SNR and image fidelity, but appears to be consistent with an edge-on dust ring with a radius of 0.3" (48 au) with a central feature as well.

\subsection{$^{12}$CO maps}
\label{sec:COlines}
\begin{figure*}[!ht]
\centering
\includegraphics[width=0.5\textwidth]{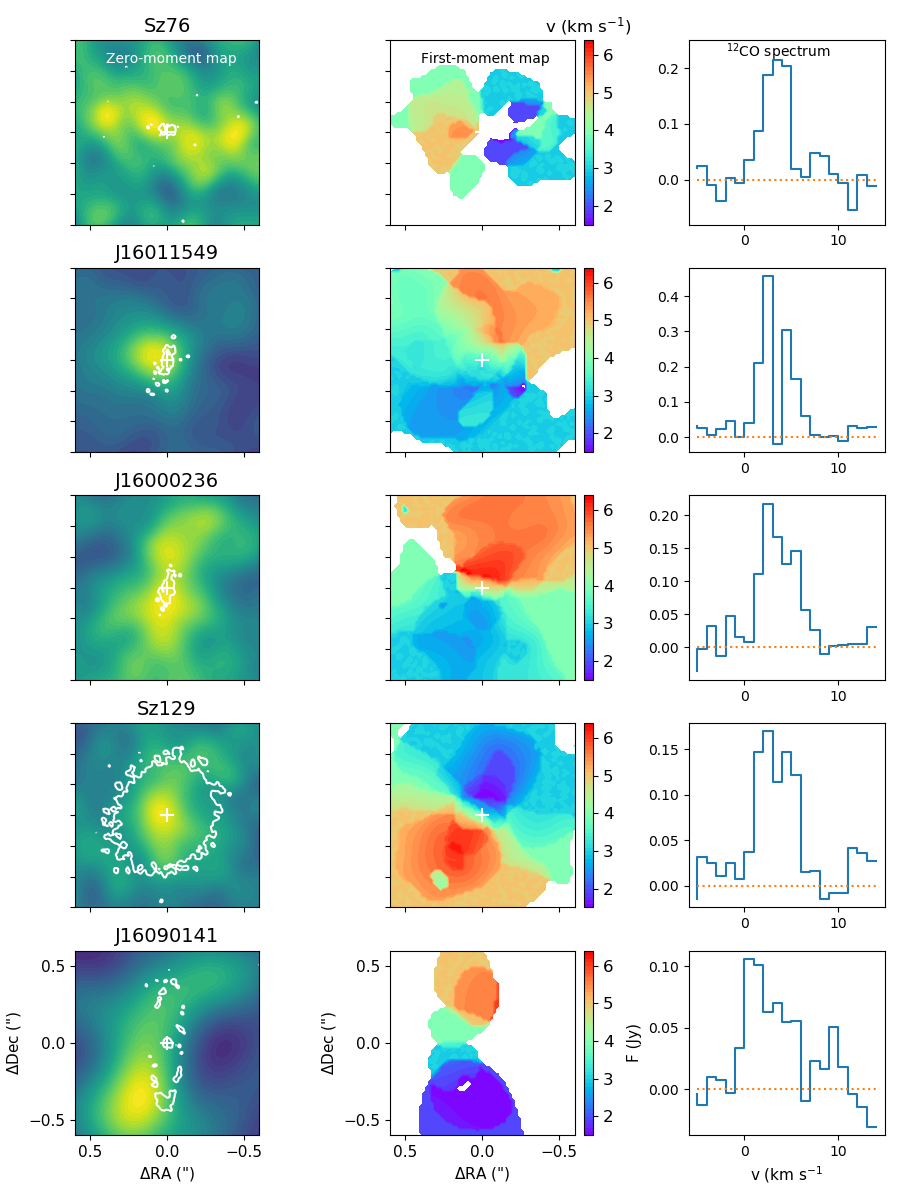}
\caption{$^{12}$CO images of the targets in our sample, where detected. From left to right the zero-moment map, first-moment map and integrated spectrum are presented, using a clipping of 3$\sigma$ in the cube. The white contours in the zero-moment map indicate the 3$\sigma_c$ threshold of the continuum images presented in Figure \ref{fig:gallery} with $\sigma_c$ the rms level.}
\label{fig:COmaps}
\end{figure*}

Figure \ref{fig:COmaps} presents the zero- and first-moment maps of the detected $^{12}$CO cubes and the integrated spectrum, with a clipping of 3$\sigma$. All CO cubes are consistent with Keplerian rotation, although J16011549 is heavily obscured by foreground absorption. CO is only detected for the 5 brightest continuum disks in the sample, with the exception of Sz103 which remains undetected despite its higher continuum flux than Sz76. This might be related to its face-on nature, as emission will be further spread out spatially. The first-moment maps (velocity maps) reveal the orientation of each disk, which appears to be consistent with the orientation as derived from the continuum emission. The SNR and image fidelity of the data are too low for detailed analysis of kinematics. 

\section{Modeling}
\label{sec:model}

\subsection{Visibilities}
\label{sec:visibility}
In order to quantify the morphology of each of these disks, the continuum visibilities of most targets are modeled using the intensity profile parametrization by \citet{Tazzari2020}:
\begin{equation}
I_{\nu} = I_0 \left(\frac{r}{r_c}\right)^{\gamma_1}\exp\left[-\left(\frac{r}{r_c}\right)^{\gamma_2}\right]
\end{equation}
Together with the position angle $PA$, inclination $i$ and offsets in RA and Dec the model contains 8 free parameters. 

The two disks with large ring-like structures (J16090141 and J16070384) are modeled instead using a Gaussian ring model plus an inner Gaussian:
\begin{equation}
I_{\nu} = I_{r}\exp\left(\frac{-(r-r_r)^2}{2r_w^2}\right) + I_{cr}\exp\left(\frac{-r^2}{2r_{wc}^2}\right)
\end{equation}
where the second term describes in the inner disk dust component. This model thus contains 9 free parameters.

The initial disk orientation and size are estimated using the \texttt{uvmodelfit} task in CASA and for J16000236, Sz129 and J16011549 initial estimates were derived using low resolution data \citep{Tazzari2020}. These parameters were used as starting point for the fitting procedure. The best-fit parameters and uncertainties are found using the MCMC fitting procedure with \texttt{galario} following \citep{Tazzari2017}, with 90 walkers and 3000-4000 steps. The last 1000 steps were used to construct the corner plots (Figures in Appendix \ref{sec:corners}) which show the posterior distributions. Table \ref{tbl:mcmc} lists the median parameters and their 68\% confidence intervals. It is important to note that the $I_c$, $r_c$ and $\gamma_1$ parameter posteriors are usually not gaussian, but show a strong degeneracy. Table \ref{tbl:mcmc} also lists the peak of each profile $R_{\rm peak}$ and the radius enclosing 68\% of the flux $R_{68}$ in au, which represent the cavity radius and outer radius, respectively.

For Sz104 and Sz112, no convergence was reached due to the large spread in $\gamma_2$, $i$ and $PA$ posteriors, as these disks could only be marginally resolved with the covered spatial scales. Instead, we run a simple Gaussian model to describe the intensity profile of these disks, following:
\begin{equation}
I_{\nu} = I_{0}\exp\left(-\left(\frac{r}{r_c}\right)^2\right)
\end{equation}

\begin{figure*}[!ht]
\centering
\includegraphics[width=1.1\textwidth,trim=100 0 0 0]{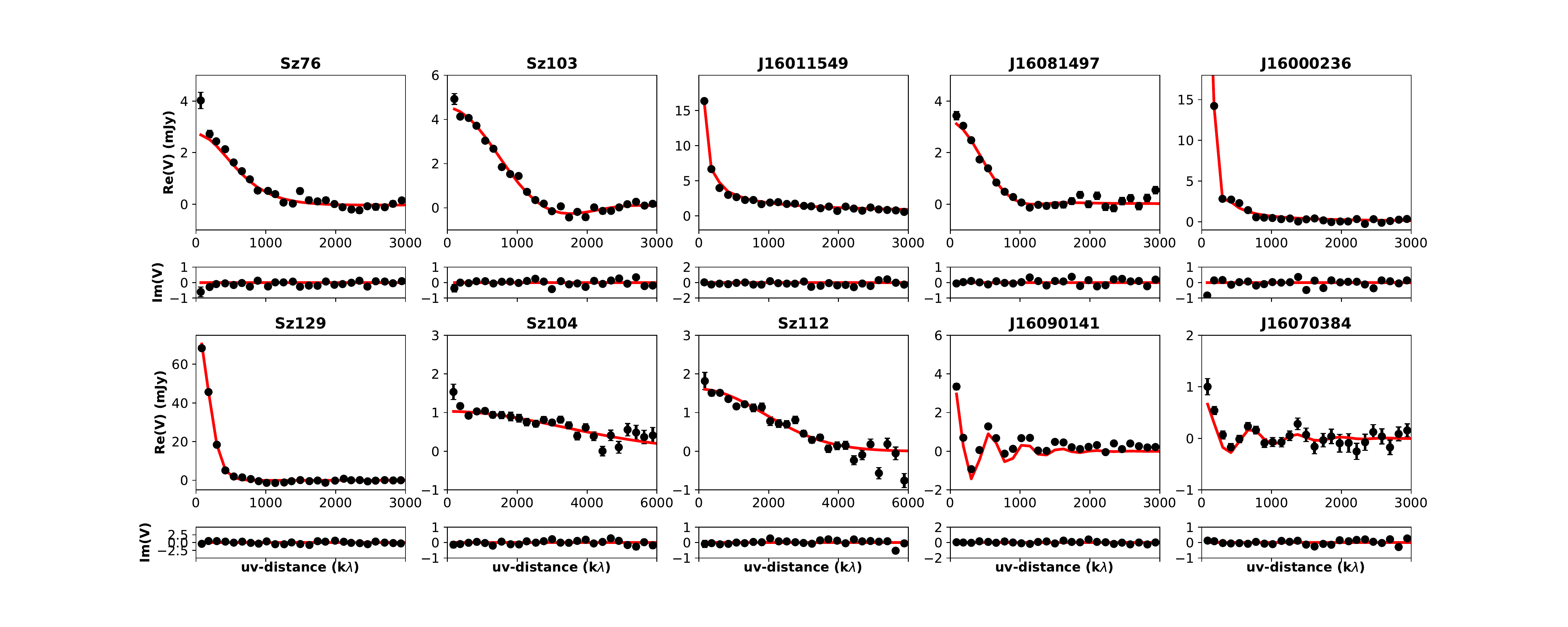}
\caption{Visibility curves of best-fit models of each of the targets.}
\label{fig:viscurves}
\end{figure*}

\begin{figure*}[!ht]
\includegraphics[width=\textwidth]{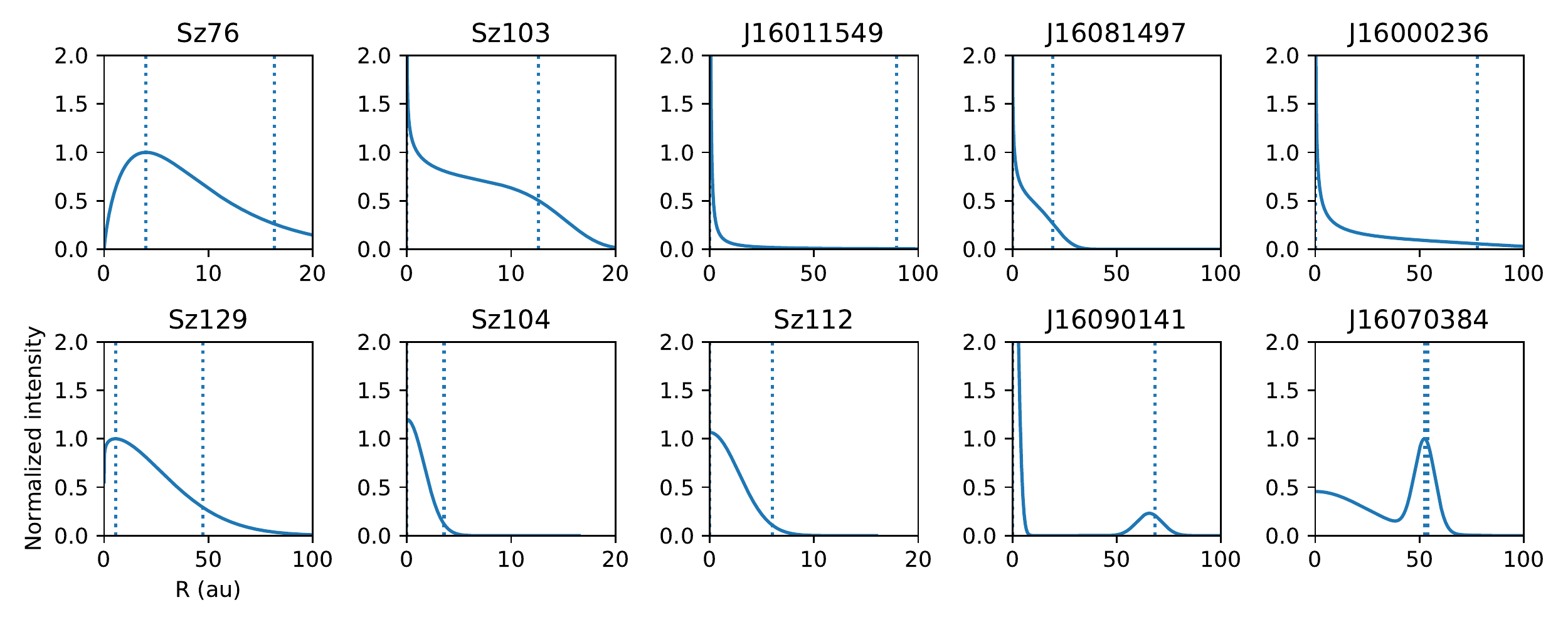}
\caption{Peak-normalized intensity profile of the best fit models of each of the targets. The dashed lines indicate the $R_{\rm peak}$ and $R_{68}$ radii which represent the cavity and outer radius.}
\label{fig:modelcurves}
\end{figure*}

\begin{figure*}[!ht]
\includegraphics[width=0.48\textwidth]{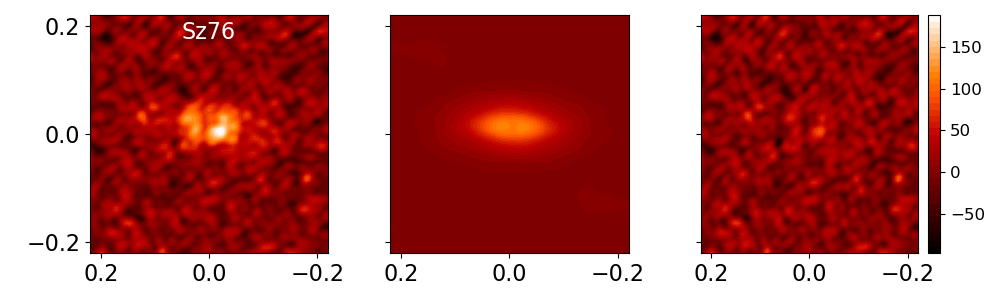}
\includegraphics[width=0.48\textwidth]{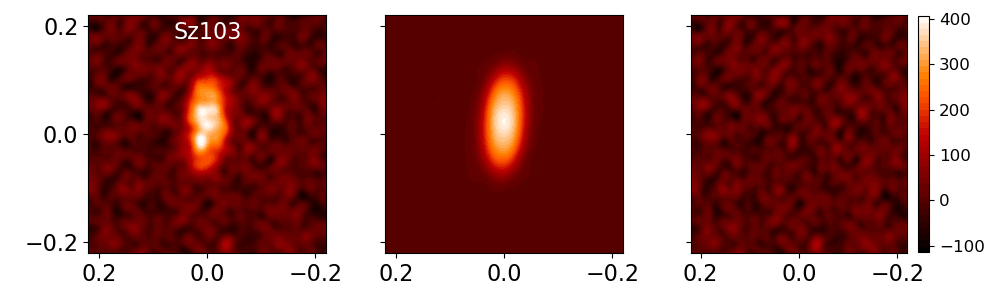}\\
\includegraphics[width=0.48\textwidth]{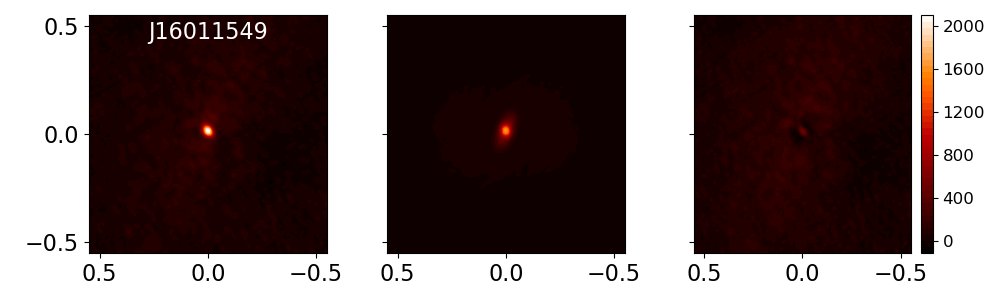}
\includegraphics[width=0.48\textwidth]{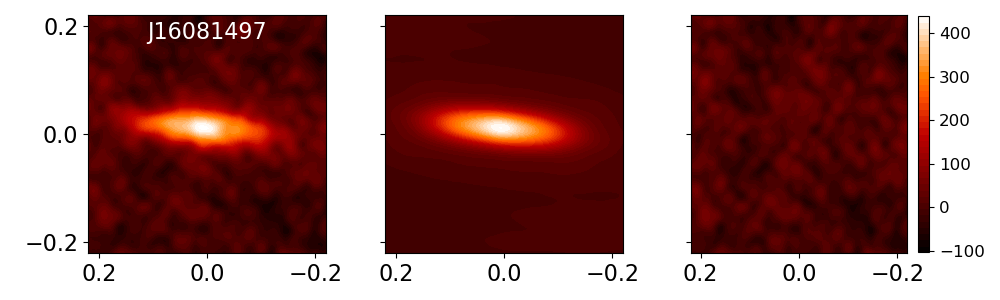}\\
\includegraphics[width=0.48\textwidth]{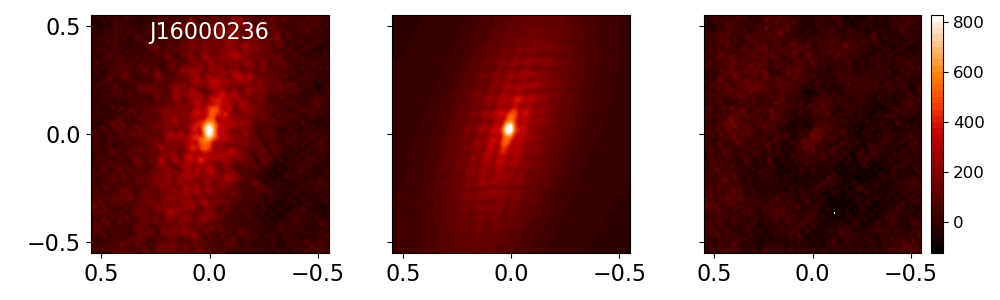}
\includegraphics[width=0.48\textwidth]{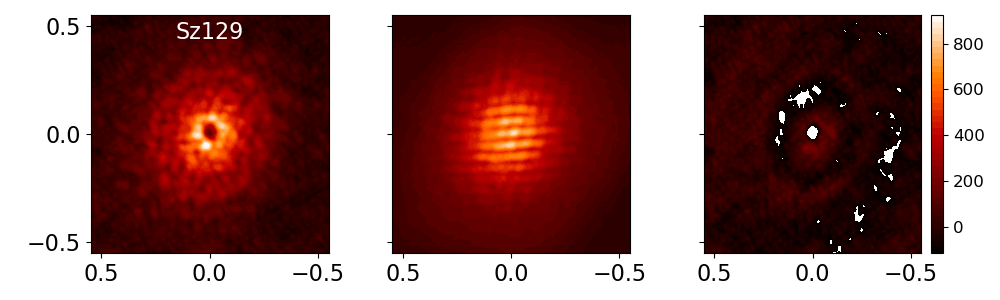}\\
\includegraphics[width=0.48\textwidth]{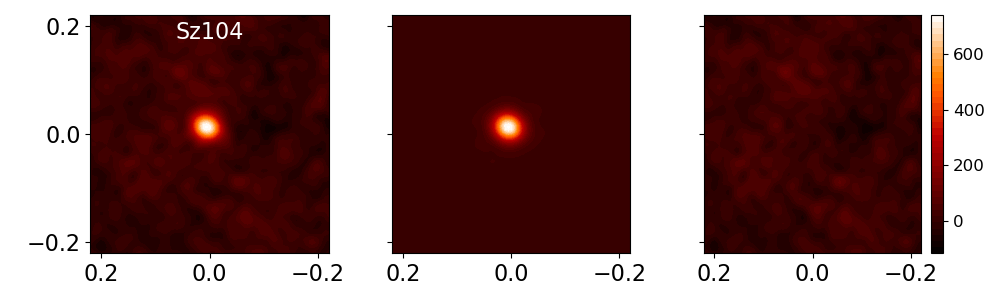}
\includegraphics[width=0.48\textwidth]{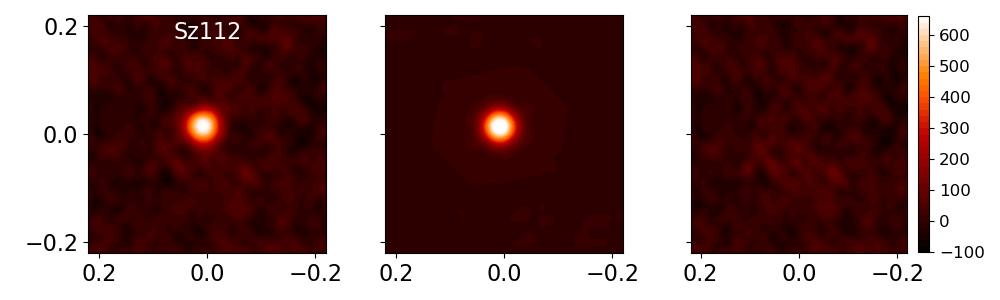}\\
\includegraphics[width=0.48\textwidth]{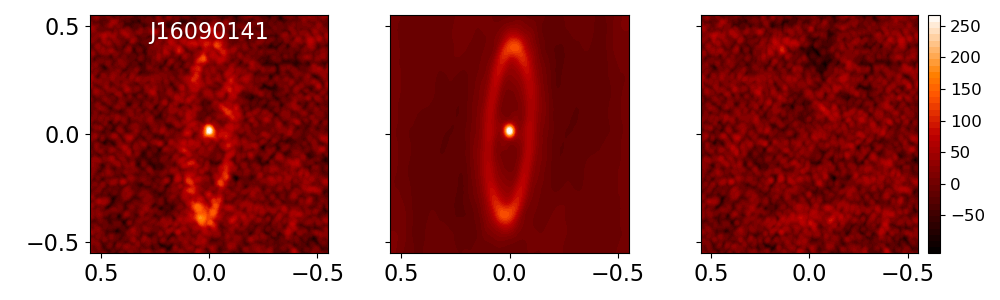}
\includegraphics[width=0.48\textwidth]{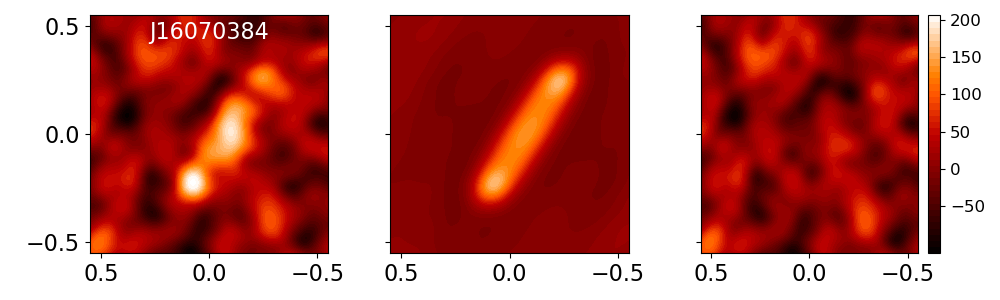}
\caption{Best fit comparison in the image plane. From left to right the original, the model (mapped on the observed visibilities) and the residual, all imaged with the same \texttt{tclean} parameters and shown on the same color scale. The flux scale (colorbar) is in $\mu$Jy beam$^{-1}$. The residual map of Sz~129 contains highly negative residuals as the fit is not representative (see text).}
\label{fig:imagecomp}
\end{figure*}

The best fit model is mapped onto the observed visibilities, subtracted and imaged using the same parameters as in the original imaging procedure. The visibility curves, model profiles and image comparisons are shown in Figure \ref{fig:viscurves}, \ref{fig:modelcurves} and \ref{fig:imagecomp}. Note that we have swapped the order to reflect the choice of the three parametric models. 

\begin{table*}[!ht]
\tiny
\caption{Visibility intensity profile fitting best-fit parameters}
\label{tbl:mcmc}
\begin{tabular}{llllllllllllll}
\hline
Name	&	$\log I_0$	&	$r_c$	&	$\gamma_1$	&	$\gamma_2$	&&	$i$	&	$PA$	&	dRA	&	dDec	   & $R_{pk}$ & $R_{68}$ & $F_{1.3mm}$\\		 
&(Jy sr$^{-1}$)&(")&&&&($\circ$)&($\circ$)&(")&(")&(au)&(au)&(mJy)\\
\hline
Sz76	&	10.44$^{+0.07}_{-0.19}$	&	0.03$^{+0.03}_{-0.02}$	&	0.81$^{+0.60}_{-0.46}$	&	1.00$^{+0.31}_{-0.23}$	&&	66.31$^{+1.24}_{-1.47}$	&	85.60$^{+1.51}_{-1.43}$	&	-0.011$^{+0.002}_{-0.002}$	&	0.019$^{+0.001}_{-0.001}$	&	4.0	&	16	&	2.7	\\ 
Sz103	&	10.19$^{+0.07}_{-0.05}$	&	0.10$^{+0.002}_{-0.002}$	&	-0.17$^{+0.10}_{-0.08}$	&	5.70$^{+2.04}_{-1.49}$	&&	66.13$^{+0.73}_{-0.74}$	&	-4.26$^{+0.90}_{-0.91}$	&	-0.005$^{+0.001}_{-0.001}$	&	0.029$^{+0.007}_{-0.007}$	&	0.0	&	13	&	4.5	\\ 
J1601	&	8.59$^{+0.01}_{-0.01}$	&	0.97$^{+0.02}_{-0.02}$	&	-1.19$^{+0.004}_{-0.004}$	&	8.92$^{+1.93}_{-1.61}$	&&	68.96$^{+0.44}_{-0.44}$	&	157.22$^{+0.52}_{-0.57}$	&	0.005$^{+0.000}_{-0.000}$	&	0.021$^{+0.000}_{-0.000}$	&	0.0	&	90	&	23.0	\\
J1608	&	9.96$^{+0.11}_{-0.06}$	&	0.15$^{+0.01}_{-0.01}$	&	-0.29$^{+0.15}_{-0.09}$	&	3.60$^{+0.91}_{-0.91}$	&&	78.29$^{+0.53}_{-0.59}$	&	84.84$^{+0.59}_{-0.57}$	&	0.002$^{+0.001}_{-0.001}$	&	0.017$^{+0.001}_{-0.001}$	&	0.0	&	19	&	3.2	\\
J1600	&	9.51$^{+0.01}_{-0.01}$	&	0.64$^{+0.004}_{-0.004}$	&	-0.59$^{+0.01}_{-0.01}$	&	3.86$^{+0.15}_{-0.14}$	&&	65.52$^{+0.19}_{-0.20}$	&	161.25$^{+0.23}_{-0.24}$	&	0.004$^{+0.001}_{-0.001}$	&	0.029$^{+0.001}_{-0.001}$	&	0.0	&	78	&	50.0	\\
Sz129	&	10.31$^{+0.02}_{-0.02}$	&	0.24$^{+0.01}_{-0.01}$	&	0.06$^{+0.02}_{-0.02}$	&	1.70$^{+0.04}_{-0.04}$	
&&	30.80$^{+0.41}_{-0.42}$	&	149.11$^{+0.78}_{-0.76}$	&	0.021$^{+0.001}_{-0.001}$	&	0.037$^{+0.001}_{-0.001}$	&	5.4$^a$	&	47	&	79.0	\\
\hline
	
	&	$\log I_0$	&	$r_c$	&	&	&	&	$i$	&	$PA$	&	dRA	&	dDec		&&&\\
&(Jy sr$^{-1}$)&(")&&&&($\circ$)&($\circ$)&(")&(")&&&\\
\hline
Sz104	&10.92$^{+0.04}_{-0.04}$	&0.014$^{+0.001}_{-0.001}$	&&&&31.00$^{+10.97}_{-16.14}$	&66.00$^{+20.12}_{-17.96}$	&0.00$^{+0.00}_{-0.00}$	&0.02$^{+0.00}_{-0.00}$ & 0.0	& 2.4 & 1.1 \\
Sz112	&10.57$^{+0.03}_{-0.02}$	&0.025$^{+0.001}_{-0.001}$	&&&&20.18$^{+10.10}_{-11.50}$	&-33.00$^{+22.03}_{-25.87}$	&0.00$^{+0.00}_{-0.00}$	&0.02$^{+0.00}_{-0.00}$ & 0.0 	& 3.4 & 1.4 \\
\hline	
	&	$\log I_r$	&	$r_r$	&	$r_w$	&$\log I_c$	&	$r_{wc}$&	$i$	&	$PA$	&	dRA	&	dDec		&&&\\
&(Jy sr$^{-1}$)&(")&(")&&(")&($\circ$)&($\circ$)&(")&(")&&&\\
\hline
J1609&	9.49$^{+0.04}_{-0.04}$	&	0.405$^{+0.003}_{-0.003}$	&	0.037$^{+0.003}_{-0.003}$	&	10.81$^{+0.11}_{-0.11}$	&	0.013$^{+0.00}_{-0.00}$ &	76.20$^{+0.31}_{-0.29}$	&	176.83$^{+0.25}_{-0.21}$	&	-0.005$^{+0.001}_{-0.001}$	&	0.020$^{+0.001}_{-0.001}$	&	64	&	68	&	4.1	\\
J1607	&	9.08$^{+0.77}_{-0.33}$	&	0.32$^{+0.02}_{-0.02}$	&	0.03$^{+0.01}_{-0.02}$	&	8.76$^{+0.24}_{-0.20}$	&	 0.15$^{+0.02}_{-0.06}$& 79.14$^{+1.64}_{-2.87}$	&	147.23$^{+1.85}_{-1.85}$	&	-0.08$^{+0.02}_{-0.01}$	&	0.07$^{+0.01}_{-0.04}$	&	52	&	54	&	1.3	\\ 
\hline
\end{tabular}
$^a$ Not a good representation of the image, see text.
\end{table*}

All fits reproduce the image morphologies, with the residuals below the 3$\sigma$ noise levels, with the exception of Sz~129: the inner cavity is not recovered and an outer ring at $\sim$0.29" radius is still present in the residual image. This implies that the model prescription is not a good representation of Sz~129. The DSHARP image of Sz129 at similar resolution but with two times higher SNR \citep{Andrews2018} also reveals the outer ring at 46 au (0.29") and even a third ring at 69 au (0.43") \citep{Huang2018}, the latter remains undetected in our image. \citet{Huang2018} derives an inner dust ring radius of $\sim$10 au.

In an alternative fitting procedure, the intensity profile is parametrized as the sum of two radially asymmetric Gaussians, following \citet{Pinilla2018tds}:
\begin{align}
I_1(r) &= C_1 \exp\left(-\left(\frac{(r-r_{ring1})^2}{2r_{w}^2}\right)\right) \\
\textrm{with } r_w&=r_{w1a} \textrm{ for } r<r_{ring1} \\
 &=r_{w1b} \textrm{ for } r>r_{ring1} \\
I_2(r) &= C_2 \exp\left(-\left(\frac{(r-r_{ring2})^2}{2r_{w}^2}\right)\right) \\
\textrm{with } r_w&=r_{w2a} \textrm{ for } r<r_{ring2} \\
 &=r_{w2b} \textrm{ for } r>r_{ring2} \\
\textrm{and }I(r) &= I_1(r)+I_2(r)
\end{align}

\begin{figure*}[!ht]
\centering
\includegraphics[width=0.45\textwidth]{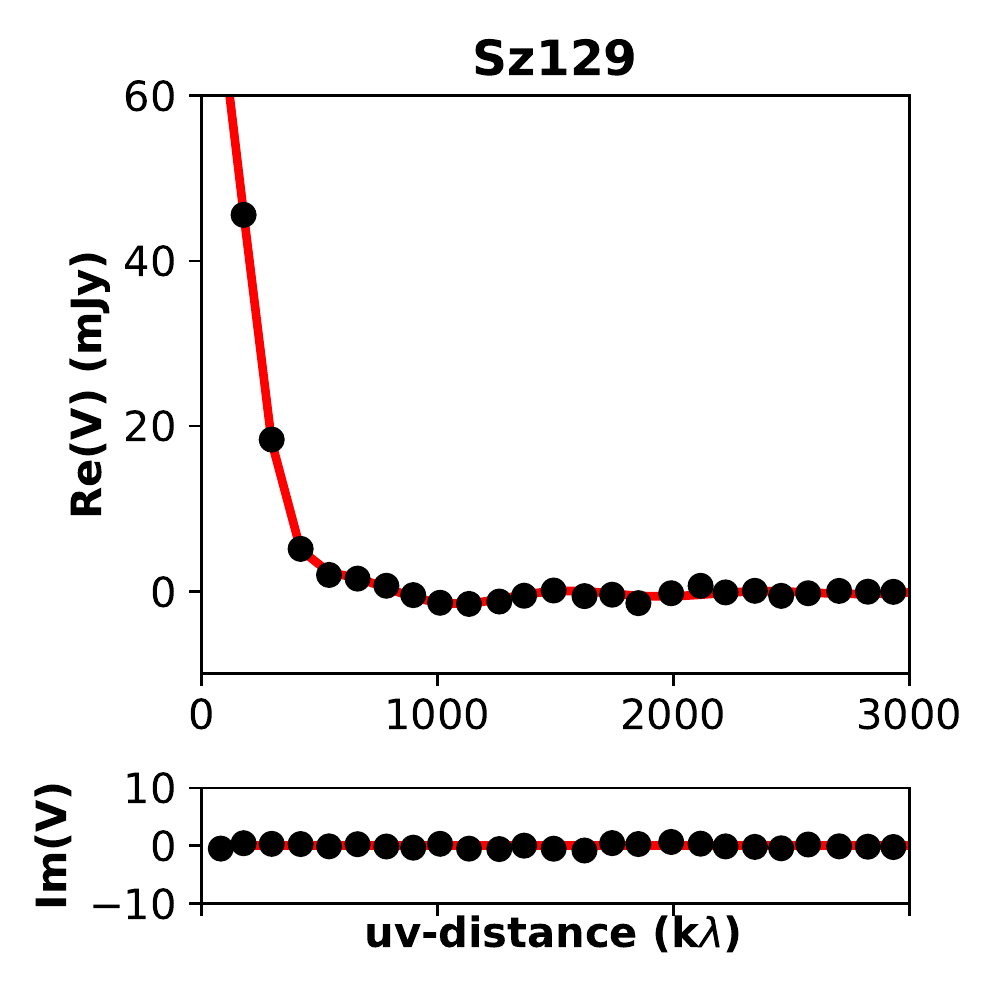}
\includegraphics[width=0.45\textwidth]{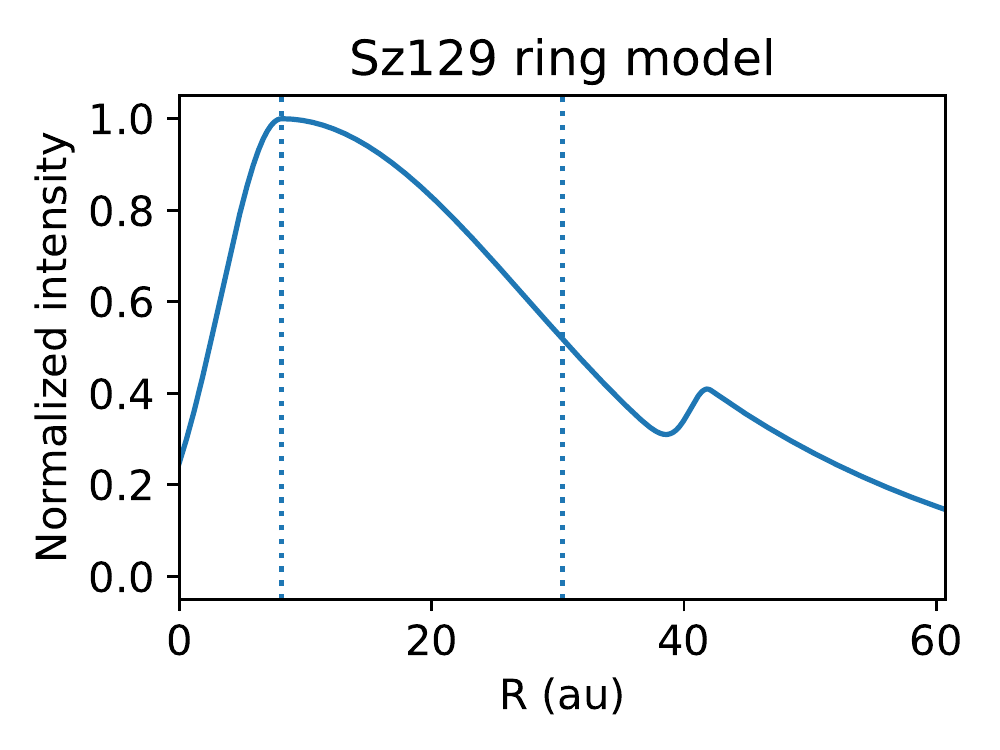}\\
\includegraphics[width=0.48\textwidth]{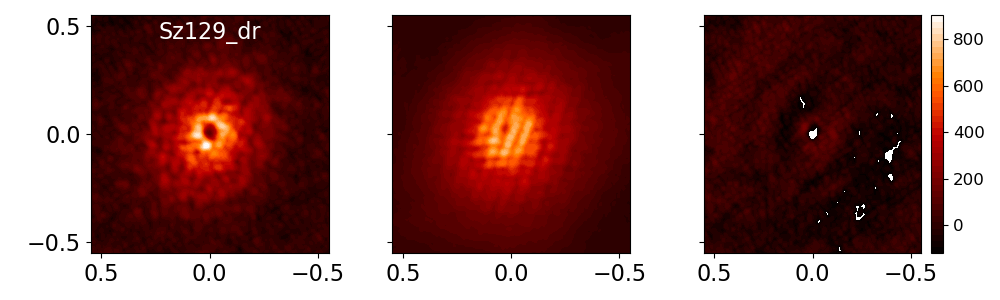}
\caption{Best fit results of the alternative double ring model for Sz129. This parametrization provides a better fit than the double power-law model presented in Figure \ref{fig:viscurves} and provides a better description of the inner cavity.}
\label{fig:sz129}
\end{figure*}

Together with inclination, position angle and RA and Dec offsets this results in a model with 12 free parameters, which is run following the same procedure as above using \texttt{galario}. The best-fit parameters are listed in Table \ref{tbl:bestfitsz129} and the visibility curve and image comparison are shown in Figure \ref{fig:sz129}. The reduced $\chi^2$ value of the best-fit of the double-ring model is lower than that of the power-law model, confirming that this is indeed a better fit. Inspection of the residual image shows that the cavity still appears somewhat too small, but the outer ring has disappeared and the residuals have improved from $\pm10\sigma$ down to $\pm$7$\sigma$. The inner cavity size ($R_{\rm peak}$ is estimated at 8.1 au, closer to the value derived by \citet{Huang2018}. The new derived values for $R_{\rm peak}$ and $R_{68}$ will be used in the subsequent analysis.

Overall, the majority of the 8 disks in the small cavity sample have an $R_{\rm peak}$ value of 0 au, which implies there is no evidence for an inner cavity in the millimeter dust emission, with the exception of Sz~76 and Sz~129 which have an inner cavity of 4 and 8 au, respectively. The two disks which could not be fit with the double power-law profile, Sz~104 and Sz~112, are well fit with a simple Gaussian profile, implicating an extremely compact disk dust radius of only 2-3 au. 

\begin{table*}[!ht]
\centering
\caption{Best fit Sz129 for double ring model}
\label{tbl:bestfitsz129}
\begin{tabular}{ll}
\hline
Parameter & Best-fit value \\
$\log C_1$ & 10.26$^{+0.005}_{-0.005}$\\
$\log C_2$ & 9.54$^{+0.02}_{-0.02}$\\
$r_{ring1}$ (") & 0.05$^{+0.005}_{-0.005}$\\
$r_{ring2}$ (") & 0.26$^{+0.006}_{-0.007}$\\
$r_{w1a}$ (") & 0.03$^{+0.007}_{-0.006}$\\
$r_{w1b}$ (") & 0.12$^{+0.004}_{-0.004}$\\
$r_{w2a}$ (") & 0.01$^{+0.007}_{-0.007}$\\
$r_{w2b}$ (") & 0.12$^{+0.004}_{-0.004}$\\
$i (^{\circ}$)& 30.9$^{+0.38}_{-0.41}$\\
$PA(^{\circ}$) & 149$^{+0.72}_{-0.71}$\\
dRA (") & 0.019$^{+0.0005}_{-0.0005}$\\
dDec (") & 0.035$^{+0.0004}_{-0.0004}$\\
$R_{\rm peak}$ (au)& 8.1\\
$R_{\rm 68}$ (au)& 51.7\\
$F_{1.3mm}$ (mJy) & 76 \\ 
\hline
\end{tabular}
\end{table*}

Both edge-on disks, J16090141 and J16070384, are consistent with a disk model with an inner disk and a large cavity size of 64 and 52 au, respectively. These two disks complement the 11 previously identified transition disks with large cavities in Lupus \citep{vanderMarel2018a}. For consistency with the analysis in this work, the Band 6 visibility continuum data of \citet{Ansdell2018} of these 11 transition disks are fit to a ring model using \texttt{galario} assuming a radially asymmetric Gaussian to derive the $R_{\rm peak}$ values, representing the cavity radii. For Sz~91, we use the Band 7 continuum data presented in \citet{Tsukagoshi2019} as it was not included in the ALMA dataset presented by \citet{Ansdell2016}. The fitting procedure and fit results are presented in Appendix \ref{sec:largecav}. Dust cavity radii of these transition disks range between 11 and 87 au. These cavity radii are used in the subsequent analysis.

\begin{figure*}[!ht]
\begin{center}
\includegraphics[width=1.1\textwidth]{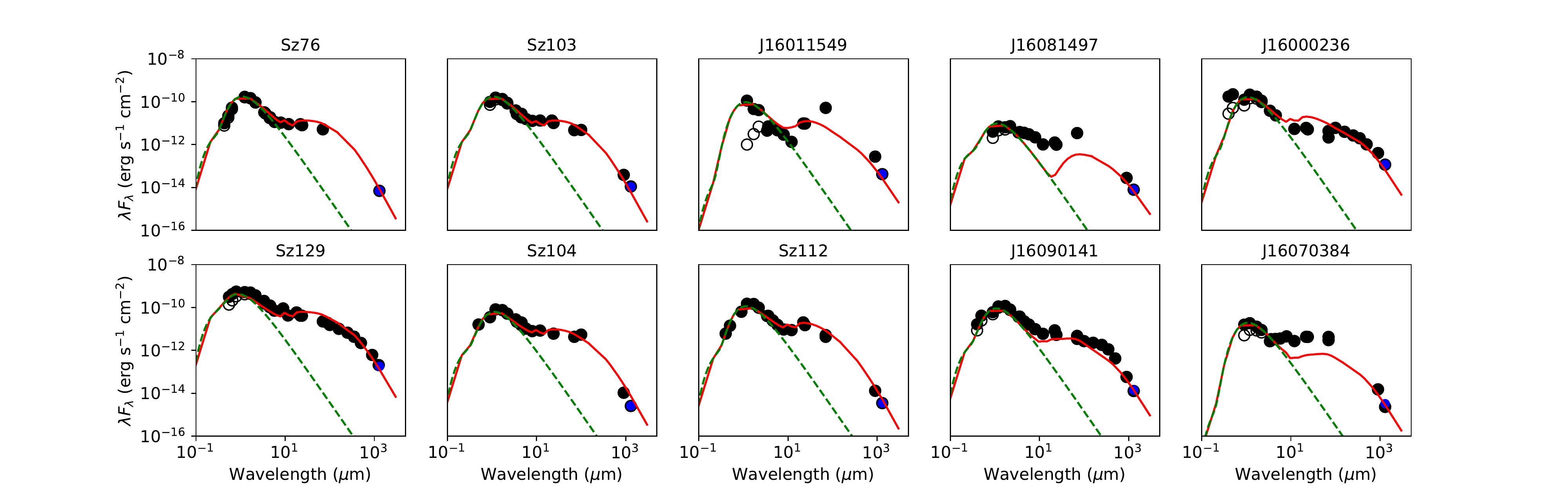}
\caption{Spectral energy distributions and radiative transfer models of all targets in our sample. The mismatch between data and model for J16081497, J16070384 and J16011549 can be understood by misalignment of the inner disk or scale height variations, as described in the text and Figure \ref{fig:sedadj}.}
\label{fig:SEDs}
\end{center}
\end{figure*}

\subsection{Radiative transfer SEDs}
\label{sec:radmc}
As six of the disks were originally selected from SED analysis, we reassess the SEDs of our targets with a radiative transfer model using the inner and outer radii as derived from the visibility analysis from the previous Section. The goal is not to fit the SEDs exactly, but to understand how the radial structure in the millimeter corresponds to features in the SED. The radiative transfer is performed using DALI \citep{Bruderer2012,Bruderer2013} which solves for the dust temperatures through continuum radiative transfer from UV to millimeter wavelengths, using a parametrized 2D dust density model and an input stellar spectrum to compute the radiation field. DALI was developed for gas surface density analysis using CO isotopologues, so the models will also be used for a comparison with the $^{12}$CO intensity profiles (see Section \ref{sec:COprofiles}). The surface density profile is assumed to be an exponential power-law, following \citep{LyndenPringle1974}:
\begin{equation}
\Sigma(r) = \Sigma_c \left(\frac{r}{r_c}\right)^{-\gamma}\exp\left(-\left(\frac{r}{r_c}\right)^{2-\gamma}\right)
\end{equation}
and is essentially set by $\Sigma_c$ and $r_c$, with $\gamma$ set at 1. The radial profile is set following the prescription in \citep{Bruderer2014}: an inner dust disk set by multiplication of $\Sigma(r)$ by $\delta_{\rm dust}=0.1$ between the sublimation radius $r_{\rm sub}$ (=0.07($L_*/L_{\odot}$) and $r_{\rm gap}$ (set to 1 au), an empty gap between $r_{\rm gap}$ and $r_{\rm cav}$ and an outer cut-off at $r_{\rm out}$=100 au (in line with the largest size of the continuum images). The vertical density profile follows a Gaussian distribution, defined by scale height $h(r)=h_c(r/r_c)^{\psi}$ with $h_c$=0.1 and $\psi$=0.15, and settling parameters $f_{ls}$ and $\chi$ of 0.2 as defined in \citet{Bruderer2014}. For a complete description of each parameter and the assumed dust properties we refer to \citet{Andrews2011}. 

In our modeling approach, the radial parameters $r_{\rm cav}$ and $r_c$ are set to match the $R_{\rm peak}$ and $R_{68}$ in the intensity profile from the visibility modeling, and the $\Sigma_c$ is adjusted in order to reproduce the total millimeter flux. For the stellar spectrum we compute a blackbody spectrum with a UV excess based on a 10,000 K blackbody with a luminosity corresponding to the on the accretion rate luminosity, using the values from Table \ref{tbl:sample}. The models are raytraced at the derived inclination of the millimeter images.

Figure \ref{fig:SEDs} shows the model SEDs overplotted on the data, where the data points originate from optical $B,V,R$, 2MASS J,H,K and \emph{Spitzer} infrared photometric data \citep{Merin2008}. The photometry has been corrected for extinction using the $A_V$ listed in Table \ref{tbl:sample}. Most models reproduce the SEDs fairly well, with the exception of J16011549, J16070384 and J16081497. The SED of J16011549 suffers from very high extinction, also seen in the $^{12}$CO emission. 
For the other two disks the infrared emission (originating from the inner disk) cannot be reproduced by our radiative transfer model. In Section \ref{sec:sedmodeling} this is further discussed.

\section{Discussion}
\label{sec:discussion}

\subsection{Constraining cavity size}
\label{sec:inclination}

\begin{figure*}[!ht]
\includegraphics[width=\textwidth]{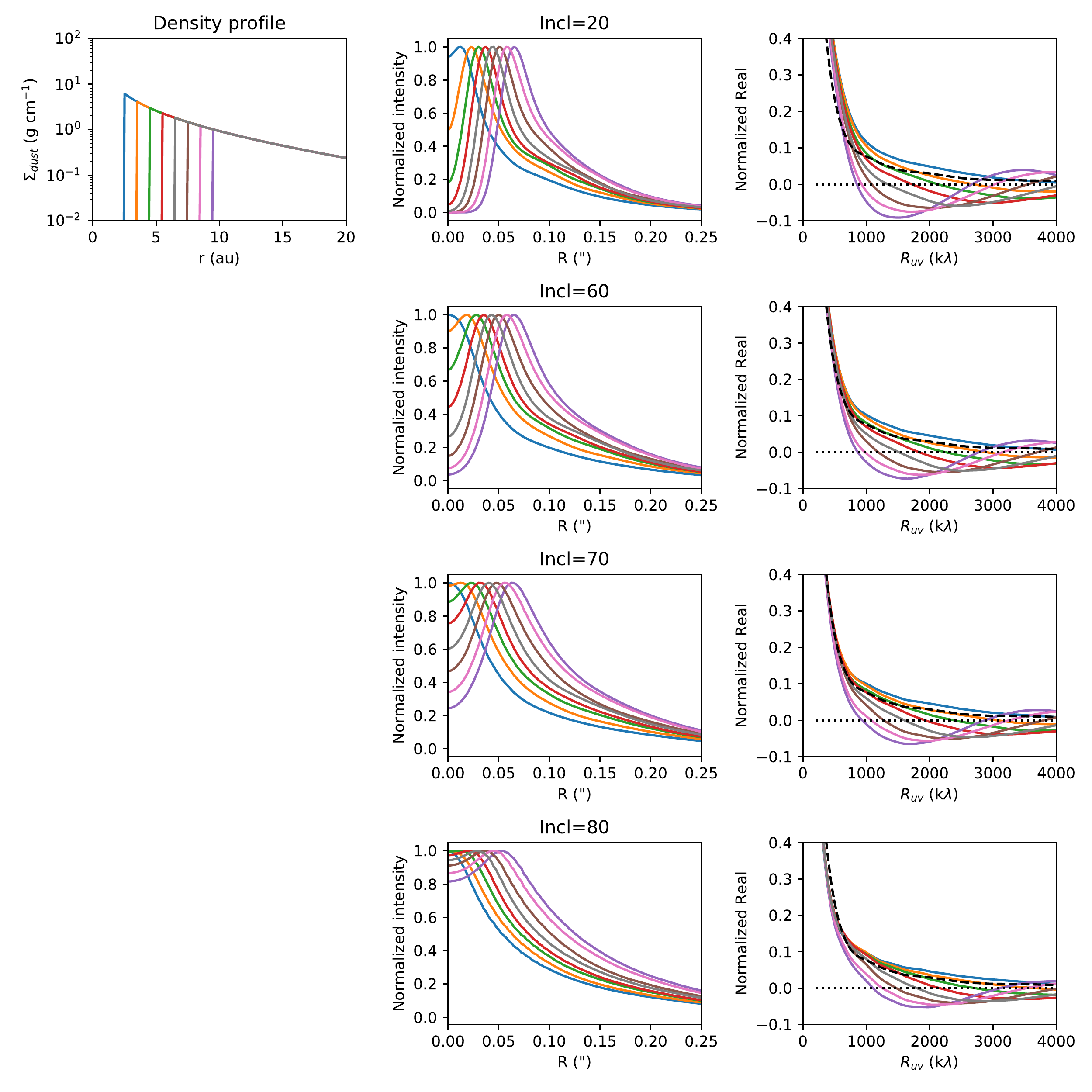}
\caption{The azimuthally averaged intensity profile and visibility curve of a series of models with a range of cavity sizes (ranging from 2-9 au) and inclinations (top to bottom: 20, 60, 70 and 80$^{\circ}$. The top left figure shows the density profile of each model, the central column the normalized intensity profile after convolution with at 0.026" beam and in the right column the normalized visibility profile. The plot demonstrates that for higher inclinations, small cavities become increasingly harder to resolve.}
\label{fig:detectability}
\end{figure*}

This study assumes that the sample of 6 transition candidates are the most promising candidates in Lupus for small inner cavities based on their SED, but the visibility analysis reveals the presence of small dust cavities with a radius of $\sim$4 au in only 1 out of 6 transition disk candidates: Sz76, and an 8 au cavity was confirmed for Sz129, hinted at from the low-resolution data \citep{Tazzari2017}. Large inner dust cavities with radii $>$50 au were found for J16090141 and J16070384. The other 6 disks are consistent with a full disk model. Considering the spatial resolution of 0.026" ($\sim$4 au diameter), this suggests that if any dust cavities are present in the other disks in the sample, the radius must be less than 2 au, assuming that the dust cavities are fully depleted of dust. However, for highly inclined disks, an inner cavity $<$10 au may remain hidden. Of the 6 disks for which the visibility analysis revealed no inner cavity, 4 disks have an inclination $>$60 degrees, the other 2 disks are extremely compact. In order to investigate the detectability of inner cavities at millimeter wavelengths at different inclinations, we run a series of generic radiative transfer models using DALI with an inner cavity size between 2 and 10 au at 20, 60, 70 and 80$^{\circ}$ inclination. The dust continuum images are convolved with a circular 0.026" beam, similar to the observations. Second, the unconvolved intensity profile are converted to a visibility curve using the analytical approach of \citet{Walsh2014hd100} to map the visibility up to 4000 k$\lambda$. Both the beam size and baseline range are comparable with our observational parameters. Figure \ref{fig:detectability} shows the comparison of the models for each inclination. The visibility curves also shows the profile for a disk without a cavity for comparison. This Figure demonstrates that for higher inclinations, small cavities become increasingly harder to resolve in both image and visibility profiles as the null shifts to larger baselines and the negative part of the Real beyond the null stays very close to zero, which becomes undetectable with typical noise levels. 

Assuming a 10\% detectability cutoff of the depth of the gap in the image (essentially a 3$\sigma$ threshold for an image with a SNR of 30), Figure \ref{fig:detectability} shows that at 20$^{\circ}$ a cavity of just over 2 au is the smallest possible scale to resolve, which becomes 3 au for 60$^{\circ}$, 4 au for 70$^{\circ}$ and 7 au for 80$^{\circ}$ inclination, respectively. In the visibility plot a threshold of 2500 k$\lambda$ for the null (beyond which most of our data become increasingly noisy) gives similar thresholds. This means that the upper limit for an inner cavity radius is $\lesssim$2.5 au for Sz104, $\lesssim$3.5 au for Sz103 and J16000236, $\lesssim$4 au for J16011549 and $\lesssim$7 au for J16081497, considering their inclinations. This sample provides the first disks with radial constraints on small, fully depleted, inner dust cavities.

Additional factors may also affect the detectability of an inner dust cavity. If dust is only partially depleted, the emission may remain optically thick and thus undetectable as an inner gap \citep{Andrews2020}. Constraining the maximum depletion is beyond the scope of this paper. It has also been suggested that SEDs show evidence for small inner cavities when only small grains have been depleted, i.e. when dust grains have grown entirely to larger sizes \citep{Dullemond2005}. Fragmentation would limit the grains from growing well beyond millimeter sizes, so a dust depletion would remain invisible at millimeter wavelengths \citep{Birnstiel2012}.

\begin{figure*}[!ht]
\centering
\includegraphics[width=\textwidth]{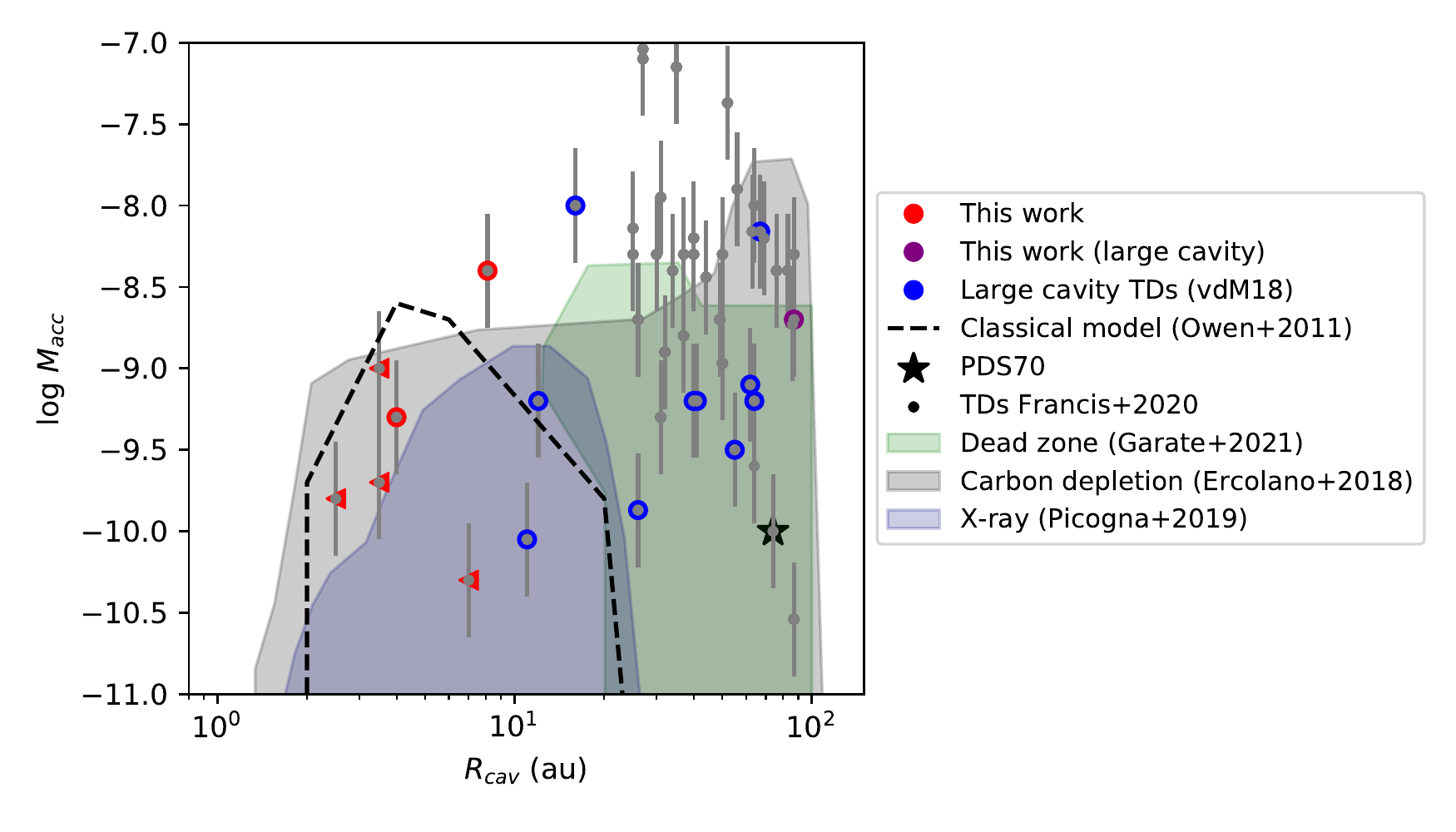}
\caption{Comparison between the derived dust cavity radii and the accretion rates of the transition disks in Lupus from this work and \citet{vanderMarel2018a}. The disks from this work where no cavity was found from the visibility analysis are indicated with a triangle as an upper limits based on their inclination, the disks with small cavities (Sz76 and Sz129) are indicated with red circles. The transition disks with large cavities from \citet{Francis2020} and the PDS70 transition disk with detected planets in the cavity are added as well. The classical photoevaporation regime from \citet{Owen2011,Ercolano2017} is shown with a dashed line, and updated photoevaporation regimes with various assumptions are shown in shaded colors \citep{Ercolano2018,Picogna2019,Garate2021}. }
\label{fig:rcavmacc}
\end{figure*}

\subsection{Origin of small cavities}
\label{sec:origin}
Transition disk cavities have traditionally been explained by two different phenomena: clearing by a (planetary) companion \citep{LinPapaloizou1979} or photoevaporative clearing \citep{Clarke2001}. 
Classical studies comparing cavity sizes and accretion rates with photoevaporation models suggest that both types of transition disks exist, where the large cavity transition disks with high accretion rates cannot be explained by photoevaporation, whereas small cavity transition disks were found to be consistent with the predictions from photoevaporative models \citep[e.g.][]{OwenClarke2012,Ercolano2017}. However, the latter study was entirely based on SED based transition disk candidates rather than confirmed cavities from millimeter continuum imaging. Furthermore, several studies with more advanced photoevaporation models have shown that larger cavity sizes with higher accretion rates could be produced as well with photoevaporative clearing  \citep{Ercolano2018,Picogna2019,Garate2021}. The new Lupus transition disk sample allows a comparison over a large range of cavity sizes and the latest photoevaporation models. 

Figure \ref{fig:rcavmacc} shows the dust cavity radii of all Lupus transition disks in comparison with their accretion rates with a typical errorbar of 0.35 dex \citep{Alcala2019}. J16070384 is excluded due to its sub-luminous nature and uncertain accretion rate \citep{Alcala2017}, and for MY~Lup we use the value from \citet{Alcala2019}. The disks without resolvable inner cavity are included as upper limits based on the analysis in Section \ref{sec:inclination}. The photoevaporation regimes from a number of modeling studies \citep{Owen2011,Ercolano2018,Picogna2019,Garate2021} overlaid. These are the regimes with a $>$0.1\% probability of finding a disk with the given cavity and accretion rate as the result of X-ray photoevaporation for different assumptions. The models by \citet{Ercolano2018} assume a depletion of gaseous carbon and oxygen of a factor of 10 in the outer disk ($>$15 au) consistent with observations \citep[e.g.][]{Miotello2017,Miotello2019}, decreasing the disk metallicity, enhancing mass loss at larger radii and creating a much wider range of inner cavities. \citet{Picogna2019} derived new prescriptions for the mass-loss rate based on 2D hydrodynamical simulations with an improved temperature structure in the wind region and found a somewhat larger regime than \citet{Owen2011} as well. Finally, \citet{Garate2021} developed photoevaporation models of disks including a dead zone in the inner part of the disk, with the explicit goal of understanding large inner cavities with high accretion rates. 

Whereas the classical photoevaporation regime by \citet{Owen2011} was only able to explain the disks with the smallest $<$10 au cavities and low accretion rates $M_{\rm acc}<10^{-9}$, these new photoevaporation regimes cover a much larger fraction of the transition disks identified from millimeter imaging in Lupus, indicating that most of them could be the result of photoevaporative clearing under the assumption of carbon depletion and/or the presence of a dead zone. The small cavity of the Sz76 disk can even be explained under the simplest assumptions of the photoevaporation model by \citet{Owen2011} and this disk is thus an ideal testbed for photoevaporative models. Whether the disks with upper limits on the cavity sizes really indicate disks with inner cavities remains an open question: these disks may have no inner cavity at all.

However, when looking at the larger known transition disk population, such as the sample of \citet{Francis2020}, which is overplotted in Figure \ref{fig:rcavmacc} as well, there is still a large number of objects that cannot be explained even with these additional assumptions. Such disks would thus still likely fall in the other category of cavities cleared by a companion. Large cavity transition disks are thought to be carved by Super-Jovian planets considering their deep gas gaps \citep{Muley2019,vanderMarel2016-isot,vanderMarel2021asym,Facchini2021}. For the smaller cavities knowledge of the gas structure is lacking, and it is unclear which planets are responsible. The pebble isolation mass  sets the minimum planet mass that halts the dust drift and creates a dust trap at its outer edge, which depends on e.g. the scale height, stellar mass and viscosity \citep[e.g.][]{Lambrechts2014,Rosotti2016,Bitsch2018,Sinclair2020,Zormpas2020}, which is $\gtrsim2 M_{\oplus}$ at 3 au for a 0.2 $M_{\odot}$ star (like Sz~76) and $\gtrsim10 M_{\oplus}$ at 20 au for a 1.0 $M_{\odot}$ star (like typical transition disk cavities). However, if the small dust is able to filter across the planet location and grow back in the inner disk, these might be only lower limits to the pebble isolation mass \citep{Drazkowska2019}.

Interestingly, the only transition disk where inner companions actually have been detected, PDS~70, actually falls in the regime that could be explained by photoevaporation as well, as shown in Figure \ref{fig:rcavmacc}. This means that the comparison of these regimes alone may not be sufficient to distinguish between the clearing scenarios. Also, the underlying assumptions of the photoevaporation models may not be applicable to all disks, for example the $\alpha$-turbulence, which remains mostly unconstrained observationally in disks \citep[e.g.][]{Flaherty2020,Tabone2021}, the origin, disk region and potentially intrinsic diversity of the carbon depletion \citep[e.g.][]{Bergin2016,Miotello2019,Krijt2020,Bosman2021}, the timescale of the gas clearing with respect to the dust evolution time scales and whether the observed dust cavity radius is actually representative for the model cavity radius \citep[see e.g. discussion in][]{Garate2021}, and the impact of the stellar mass on the mass loss rate \citep{Picogna2021}. In particular, all models were run for a 0.7 $M_{\odot}$ star whereas our sample contains a range of stellar masses.

The fraction of transition disks consistent with photoevaporative models with respect to the total disk population is generally used to estimate the dispersal timescale as the result of photoevaporation. Using the classical Owen model and a sample of SED-selected transition disks, this dispersal timescale was estimated to be $\sim10^5$ years \citep{Ercolano2017}. Our study has shown that SED-selected transition disks are in fact biased towards inclined disks which may not contain a cavity at all, which would lower this fraction, whereas transition disks without an indicator in the SED may exist as well, which would increase the fraction. The different photoevaporation models result in a different amount of transition disks as well. Regardless of the assumed model, it is clear that deriving a dispersal timescale from SED-selected transition disks remains highly debatable and high-resolution imaging of all Class II disks in Lupus is required to derive a true fraction.

\begin{figure*}[!ht]
\includegraphics[width=\textwidth]{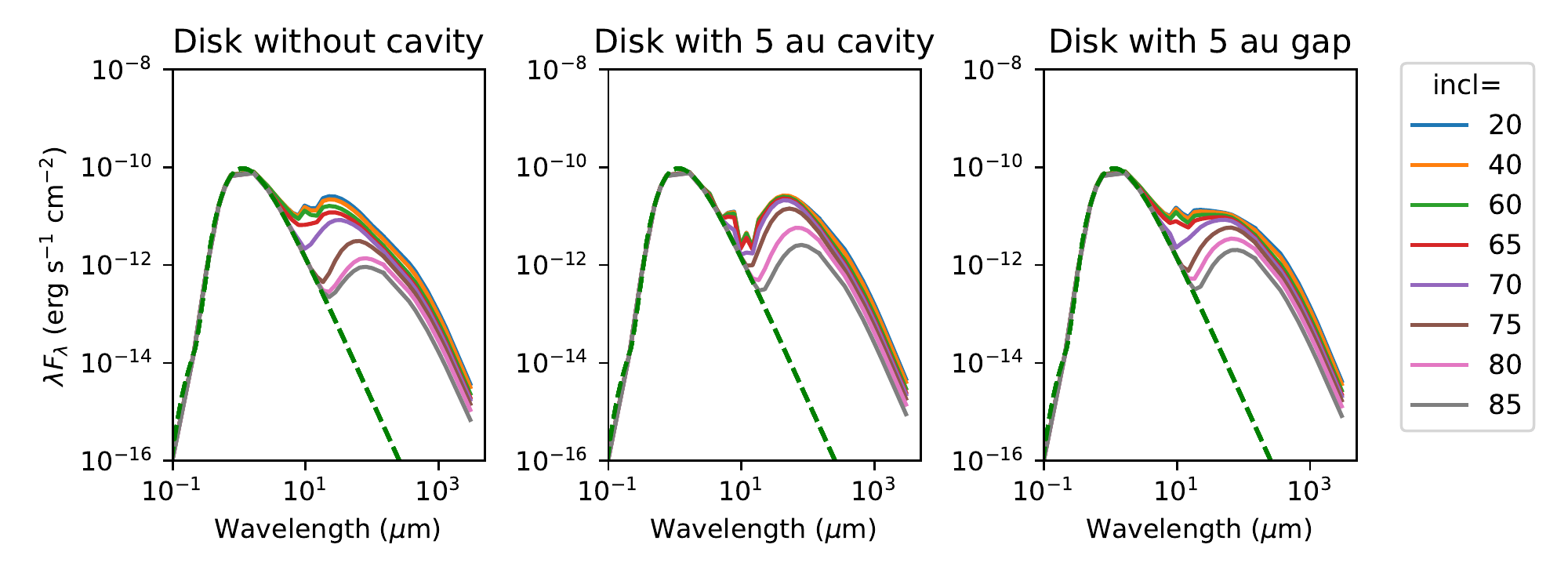}
\caption{Generic radiative transfer models for a disk without inner gaps (left), a 5 au cavity without inner disk (middle) and a 5 au gap with an inner disk (right) for a number of inclinations (see legend). Both inclination and the presence of an inner disk have a pronounced effect on the infrared emission, which could be mistaken for the presence of an inner cavity.}
\label{fig:sedvar}
\end{figure*}

\begin{figure*}[!ht]
\includegraphics[width=\textwidth]{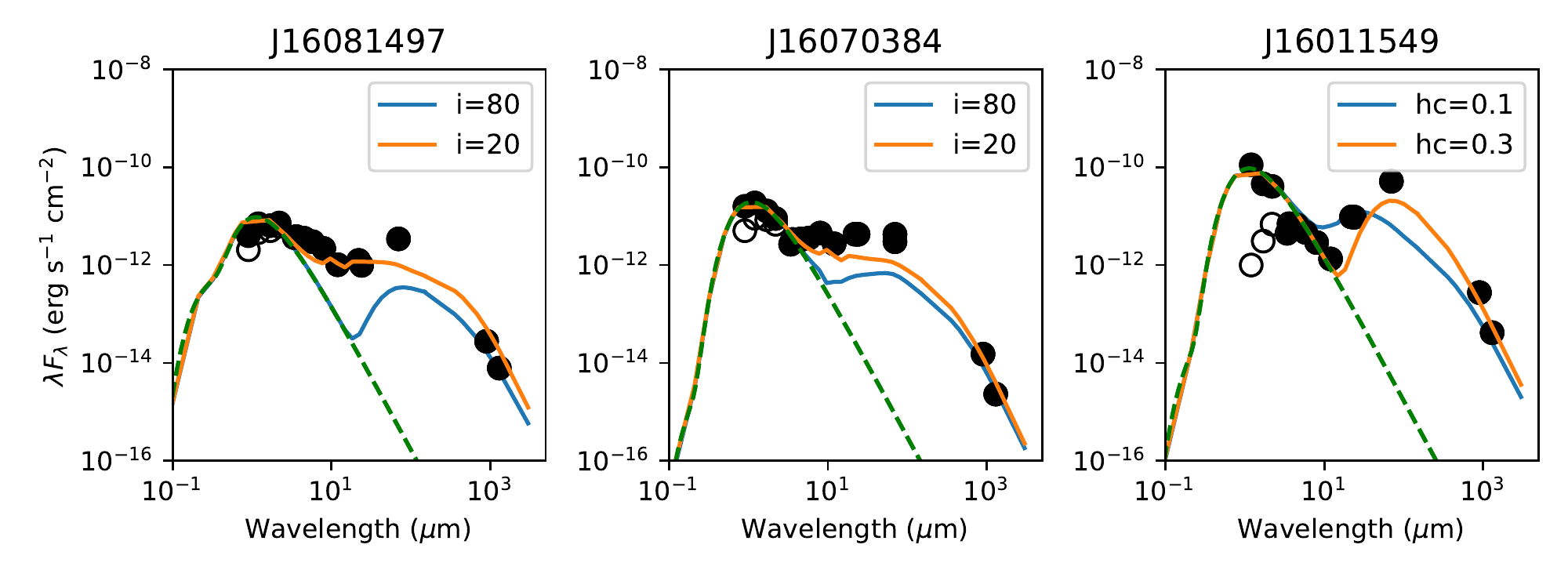}
\caption{Variations in models for J16081497, J16070384 and J16011549 (from left to right). The left two panels show that a lower inclination than the high inclination derived from the millimeter image can readily reproduce the near infrared excess observed in the SED, suggesting that the inner and outer disk may be misaligned. The right panel shows that an increase in the scale height can reproduce the high far infrared excess observed in J16011549 without changing the radial structure.}
\label{fig:sedadj}
\end{figure*}

\subsection{SEDs of transition disk candidates}
\label{sec:sedmodeling} 

This study shows that SED-selected transition disk candidates with small cavities do not necessarily show cavities in the millimeter distribution. The inclination plays an important role here: all SED-selected disks except Sz104 and Sz112 have inclinations of 65$^{\circ}$ and above, which limits the detectability of an inner disk cavity to several au (Section \ref{sec:inclination}) but a high inclination may even affect the appearance of the SED, in particular an apparent dip in the infrared emission which can be mistaken as a deficit of dust material \citep{Merin2010}. However, obtaining the disk orientation from the SED directly is very challenging, so SED modeling works usually start from the assumption that the majority of disks is not inclined and can thus be compared with moderately inclined disk models. The presence or absence of an inner dust disk will also have a pronounced effect on the infrared emission in the SED. Sz129, J16090141 and J16070384 and a number of disks from the large cavity study in Lupus \citep{vanderMarel2018a} have a large dust cavity but no apparent dip in their infrared emission, likely due to the presence of an inner dust disk. 

In order to understand the influence of inclination and an inner dust disk, a number of generic radiative transfer models using the basic properties of J16011549 have been run, for a range of inclinations and radial disk structures. In particular, we show the differences between a disk without cavity, a 5 au gap (with an inner dust disk) and a 5 au cavity (without inner dust disk) for a range of inclinations. The results are shown in Figure \ref{fig:sedvar}.

This Figure reveals that even in a full disk, a deep infrared dip may be seen in the SED if the inclination is high ($>65^{\circ}$). The difference between a full disk and a gapped disk with an inner dust disk is marginal. Only in the absence of an inner dust disk can a 5 au cavity be identified regardless of the inclination. This behaviour can be explained as the emitting surface of a disk becomes very small for high inclinations, which can result in the complete disappearance of infrared excess. This may lead to misidentification of transition disk candidates from SEDs. Only millimeter imaging can reveal the inclination of transition disk candidates and the possible deficit of dust material close to the star.

This behaviour may also explain the appearance of the SEDs of J16081497 and J16070384, which are both showing near infrared excess which cannot be reproduced in our radiative transfer model with the inclination as set by the outer disk image ($\sim80^{\circ}$). However, if the inner disk is misaligned with the outer disk (a `warp'), its inclination would be lower and near infrared excess is produced. This is demonstrated in the left two panels of Figure \ref{fig:sedadj}. This signature in the SED as the result of misalignment has been suggested before for the RY~Lup disk \citep{vanderMarel2018a,Bohn2021}, which also shows a high near infrared excess in combination with a strongly inclined outer disk with a large inner cavity, just like J16070384. The exact inclination of the inner disk cannot be constrained, as shown in Figure \ref{fig:sedvar}: the SED is indistinguishable for any inclination angle $\lesssim65^{\circ}$. Interestingly J16081497 does not show an inner gap in the millimeter image, so the misalignment may be more gradual in that disk. On the other hand, with a 78$^{\circ}$ inclination a small inner dust gap may simply remain hidden in the millimeter image as well. Regardless, the SED indicator of a dust gap is likely just the result of the high inclination in this case. We also note that three other highly inclined extended disks exists in Lupus \citep[Sz133,Sz69,Sz65][]{Tazzari2018} which have not been identified as transition disk candidate in the literature from their SED, so the effect of inclination may still vary from disk to disk. Generally, inclination may have a profound effect on the detectability of structures in both SED and millimeter images. Based on this work, inclination appears to have a more significant impact than the presence of an inner cavity. Considering the non-uniformity of the spatial resolution of the Lupus disk millimeter images and the lack of a complete SED study of the Lupus disks, it is not possible to derive any statistical conclusions on the expected impact of inclination on transition disk occurrence at this stage.

Misalignments between inner and outer disks are commonly observed in transition disks, either through shadows \citep{Marino2015}, CO warps \citep{Loomis2017}, optical dippers \citep{Ansdell2020}, VLTI observations \citep{Bohn2021} or directly from millimeter observations resolving the inner dust disk \citep{Francis2020}. Such a misalignment is thought to be caused by the presence of a companion which breaks the disk, resulting in a precession of the inner disk w.r.t. the outer disk \citep{Facchini2017warps,Zhu2019}. The SED indicator in combination with the orientation from the millimeter image is an additional method to reveal misalignments in inclined protoplanetary disks. 

Overall, we can conclude that SED selection criteria to identify transition disk candidates are only partially reliable. For large cavity transition disks (11 from \citet{vanderMarel2018a} and 2 from this work), only 7 had been identified before as transition disk candidate \citep{Merin2008,vanderMarel2016-spitzer}, while 6 were identified from the millimeter image only. For small cavity transition disks, only 1 out of 6 SED candidates revealed an inner cavity in the millimeter image, whereas 1 disk (Sz129) revealed a 8 au cavity without an indicator in the SED. Three out of the five transition disk SED-candidates are highly inclined. The other two disks, Sz104 and Sz112, are very compact and do not reveal an inner cavity in the millimeter-dust continuum. As their SED shows a dip, this may be the result of increased grain growth in the inner disk, effectively removing the $\mu$m-sized grains, but not the millimeter-sized grains \citep{Birnstiel2012}. Increased grain growth in the inner disk is expected when most of the dust of the original disk has drifted inwards.

\begin{figure*}[!ht]
\centering
\includegraphics[width=\textwidth]{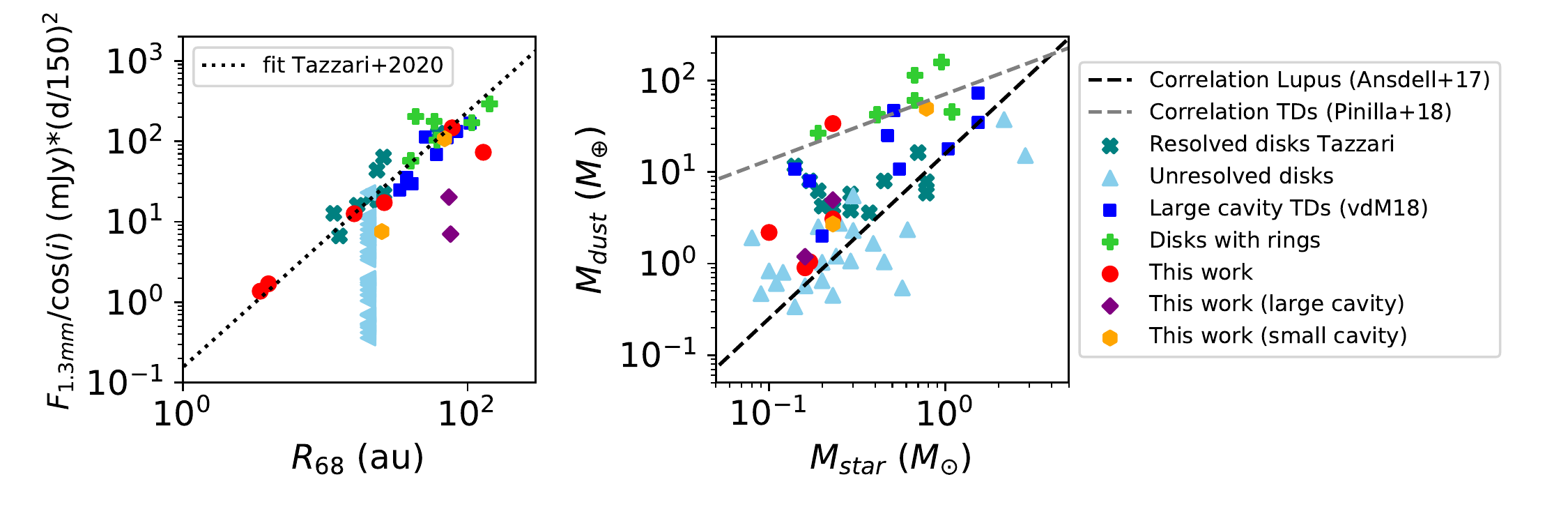}
\caption{{\bf Left:}Inclination corrected size-luminosity relation of Lupus disks, based on the work of this study and with the Band 6 results from \citet{Tazzari2020} for the low resolution data. Unresolved disks from \citet{Ansdell2018} with an upper limit on the radius of 20 au are included. The disks with inner cavity (based on \citet{vanderMarel2018a} and this work) are marked in blue and the disks that show ring structure according to \citet{Huang2018} in orange. The dotted line indicates the best fit of the size-luminosity relation as derived by \citet{Tazzari2020}. Whereas the disks from the Tazzari sample cannot be confirmed to be in the drift-dominated regime, the two compact disks from our work (Sz104 and Sz112) are following the predictions of a drift-dominated evolution model \citep{Rosotti2019}. {\bf Right:} Correlation between $M_{\rm star}$ and $M_{\rm dust}$ of all disks in Lupus, categorized as either disk with inner cavity (as defined in this work), a ring disk (as resolved by DSHARP), resolved disks (as identified by \citet{Tazzari2020}) and unresolved disks (remaining disks from \citet{Ansdell2018}). Parameter values are taken from \citet{Manara2018}, which correct for the Gaia DR2 distance. The correlation as fit by \citet{Ansdell2017} for the entire Lupus population and the correlation for transition disks as fit by \citet{Pinilla2018tds} are overlaid. There is no obvious correspondence between disks with and without substructure.}
\label{fig:sizelum}
\end{figure*}

Based on this study, the number of transition disks from SED modeling may be underestimated by at least a factor of 2, but this factor remains highly debatable due to the small sample statistics and the lack of high resolution images of the bulk of the Lupus protoplanetary disks, as many more cavities such as seen in Sz129 (without infrared dip) may still have to be revealed. Second, dust gaps in the inner part of the disk may remain invisible if they are not sufficiently deep to become optically thin as the surface density in the inner 10 au of the disk is generally so high that most continuum emission is optically thick \citep{Andrews2020}. High resolution images with the ngVLA will further enhance our knowledge of small inner disk gaps.

\subsection{Dust evolution relations}
\label{sec:sizelum}
The derived parameters and classifications can provide new insights on dust evolution in disks. This disk survey is the first high-resolution disk survey of faint protoplanetary disks, in contrast to for example the DSHARP survey which focused on the brightest protoplanetary disks. As dust continuum images of protoplanetary disks generally follow the size-luminosity relation \citep{Tripathi2017,Andrews2018size}, fainter disks are thought to be smaller in size. The size-luminosity relation has been constrained down to $\sim$20 au radius \citep{Hendler2020,Tazzari2020,Sanchis2021} for fluxes $\gtrsim$10 mJy. \citet{Long2019} analyzed the size-luminosity relation as well for a sample of disks in Taurus, including a couple of faint disks $<$10 mJy. As our sample consists of several disks in the 1-5 mJy range for which the structure is resolved, this means that the size-luminosity relation can be explored further down to check if the relation is still valid, in particular for the two very compact disks, Sz104 and Sz112.

Figure \ref{fig:sizelum} shows the size-luminosity relation based on the disks in this study and the Band 6 results from the analysis by \citet{Tazzari2020}, with a correction for inclination. This correction is important as the emitting surface decreases for higher inclinations, which are over-abundant in our sample, which means that only two disks are well below the regime traced by \citet{Tazzari2020}. The disks with inner cavities (both small and large) are indicated and the unresolved disks from \citet{Ansdell2018} are included as upper limits with a radius of 20 au.

\citet{Tazzari2020} fit the relation at Band 6 using the following equation:
\begin{equation}
\log(R_{68}({\rm au})) = A + B\log\left[\frac{F_{\nu}{\rm (Jy)}}{\cos i}\left(\frac{d}{150 pc}\right)^2\right]
\end{equation}
with best fit parameters $A$=2.4$\pm$0.1 and $B$=0.63$\pm$0.08 which is overlaid in Figure \ref{fig:sizelum}. Modeling papers tend to reverse this relation as $L_{mm}\propto R_{eff}^q$, with $q$=2 for drift-dominated disks, $q$=1 for fragmentation-dominated disks and $q$=5/4 for disks with strong substructures where the bulk of the dust mass is in the  optically thick dust rings \citep[][]{Rosotti2019,Zormpas2022}. The derived fit by \citet{Tazzari2020} of the Band 6 data corresponds to $q$=1.6$\pm$0.2, which remains inconclusive about which mechanism dominates the dust evolution. However, the results from their multi-wavelength analysis is inconsistent with predictions from radial drift and the relation can be explained by a population of disks with unresolved structure.

\begin{figure*}[!ht]
\centering
\includegraphics[width=\textwidth]{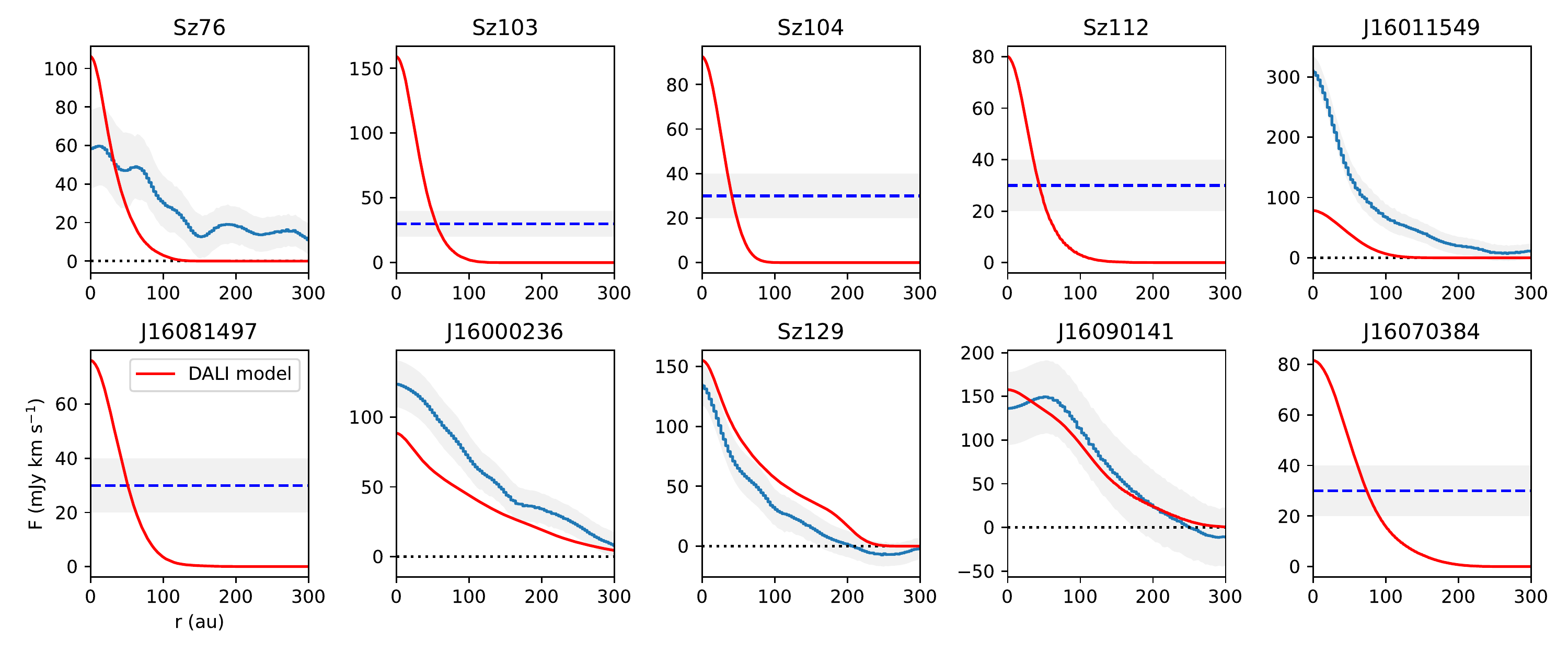}
\caption{Intensity profile of the $^{12}$CO 2--1 emission of the disks in our sample, with the data in blue and the model prediction from our DALI models in red. For the 5 disks that were undetected in $^{12}$CO, a 3$\sigma$ threshold is included as a blue dashed line. The plots demonstrate that the observed $^{12}$CO emission is  consistent with the predictions from the DALI models, including the non-detections.}
\label{fig:12codali}
\end{figure*}

Figure \ref{fig:sizelum} shows that more than half of the disks in the Tazzari sample are transition disks with large inner cavities or disks with ring-like structure \citep[as resolved by][]{Huang2018}. This is consistent with their conclusion that the disks do not follow the predictions of drift-dominated evolution. The main issue is that for $R_{eff}\gtrsim$30 au, the slope $q$ of the $L_{mm}\propto R_{eff}^q$ relation is hard to determine from the data to distinguish between the different types of dust evolution \citep[][]{Rosotti2019,Zormpas2022}. Structured disks and fragmentation dominated disks are expected to lie well above the relation in the $<$10 au regime, but such disks have not been observed yet. Interestingly, the two compact disks are shown to lie right on top of the relation computed by Tazzari. As their derived slope is in between that of drift-dominated and structure-dominated disks, it could indicate that the population contains a mix of the two disks, and considering their location, these two disks are truly drift-dominated disks and do not contain unresolved substructure. 
The other disks of our sample with low observed flux($<$5 mJy) are shifted towards higher values due to their high inclinations. Noticeable are the two ring disks J16090141 and J16070384 which are well below the derived correlation, i.e. they have a low flux for their size. These are both large scale edge-on ring disks which are essentially more like large cavity transition disks, and exceptions to the general idea that large cavity transition disks are amongst the brightest disks in the disk population \citep{OwenClarke2012,vanderMarel2018a}. Perhaps these disks are already more evolved and the dust grains have already grown to planetesimals in their pressure bumps, lowering their millimeter flux \citep[][]{Stammler2019,Pinilla2020}. Alternatively, they could be the result of lower initial disk masses \citep{Zormpas2022}.

The size-luminosity diagram thus indicates that the two compact disks from our sample could be drift-dominated, following the predictions from dust evolution models. An expansion of high resolution observations of the Lupus disks, in particular the unresolved ones, is required to further explore this scenario. Also deeper ALMA observations of the $^{12}$CO emission of these compact disks to determine the gas/dust disk size ratio will shed light on the question if they are truly drift-dominated or if the disks are intrinsically smaller.

Second, we investigate the $M_{\rm star}-M_{\rm dust}$ correlation in Lupus using the newly identified transition disks from this work and the ring disks as identified by DSHARP \citep{Huang2018}. The right panel of Figure \ref{fig:rcavmacc} shows the correlation, using the stellar and disk parameters from \citet{Manara2018} which have been corrected for the Gaia DR2 distance. The transition disks and ring disks are labeled accordingly, and the remaining disks are labeled as either resolved (as identified by \citet{Tazzari2020}) or unresolved (remaining disks). The best-fit correlation derived by \citet{Ansdell2017} for the entire Lupus population and the best-fit correlation derived by \citet{Pinilla2018tds} for a transition disk population are overlaid as well, although this transition disk sample is biased towards the brightest disks. Transition disks with large inner cavities $>$20 au have been suggested to follow a flatter correlation due to dust evolution, as their pressure bumps limit the radial drift \citep{Pinilla2018tds,Pinilla2020,vanderMarelMulders2021}. However, our plot is inconclusive about this result, as the new transition disks (at least those around the low-mass stars) actually do follow the general Lupus correlation for full disks. One possible explanation is that the dust grains in these disks have already  grown to boulders which is indeed more efficient around low-mass stars \citep{Pinilla2020}, decreasing the observable millimeter-dust mass. The bulk of the transition and ring disks around $>$0.5$M_{\odot}$ stars are consistent with the correlation derived by \citet{Pinilla2018tds}, suggesting efficient trapping in these disks. The unresolved disks do follow the general trend, consistent with disks dominated by radial drift. The resolved disks by \citet{Tazzari2020} lie in between the two correlations, suggesting they may include both disks with and without substructure. 

Sz104 and Sz112 are the most compact dust disks resolved with ALMA to date, with effective radii of only 2.5 and 3.5 au. Only a handful of other compact dust disks $<$15 au around single stars have been spatially resolved so far: CX~Tau \citep[14 au][]{Facchini2019}, CIDA~7/J0420/J0415 \citep[9.3, 15 and 10 au, respectively][]{Kurtovic2021}, and GK~Tau \citep[8.4 au][]{Long2019}. Resolved compact dust disks in binary systems also have sizes $<$15 au \citep{Akeson2019,Manara2019,Rota2022}. Also the majority of the gas disks has been suggested to be compact \citep{Miotello2021}. However, the majority of disks of the ALMA disk surveys in nearby star forming remain unresolved at $\sim$20 au radius resolution, suggesting the existence of many more compact disks. In fact, \citet{vanderMarelMulders2021} suggest that if all compact disks $<$40 au radius are drift-dominated, they could be linked directly to the systems that are likely to contain super-Earths through their stellar mass dependence \citep{Mulders2021b}. The existence of compact disks has also been suggested in connection with inward transport of H$_2$O and CO ice as the result of pebble drift \citep{Banzatti2020,Kalyaan2021,vanderMarel2021-c2h}. ALMA observations of protoplanetary disks have so far focused primarily on bright, extended disks which usually contain substructure, and it is clear that a better understanding of the dust evolution processes requires high-resolution studies of the faint, compact disks instead.

\subsection{Gas vs dust}
\label{sec:COprofiles}
The distribution of the gas compared to the dust, in particular the gas and dust radii, can provide further information on the main processes in the disk, in particular the efficiency of radial drift and indications of gas gaps. Unfortunately the $^{12}$CO emission was only detected in 5 of the disks in our sample (Section \ref{sec:COlines}) and suffers from foreground absorption and lack of large spatial scales, so no quantitative information could be derived from the moment maps. 

The question is whether the lack of detected $^{12}$CO emission in the other 5 disks can be understood, as $^{12}$CO emission tends to be bright down to low gas masses. The DALI models fit to the SEDs from Section \ref{sec:radmc} are raytraced in $^{12}$CO and the output of the line cubes is convolved, integrated and compared with the observed line emission in Figure \ref{fig:12codali}. The observed $^{12}$CO emission is consistent with the predictions from the DALI models: the detected emission is comparable in intensity, whereas the non-detections would require much more sensitive observations to detect the $^{12}$CO in the outer disk, where the bulk of the emission is emitted. However, the bright central $^{12}$CO peak from the models remains undetected, suggesting that the gas temperature of the inner part of the disk may be lower than predicted.

\section{Conclusions}
\label{sec:conclusions}
In this work we analyze high-resolution ALMA continuum images of 10 transition disk candidates from the Lupus star forming region. Six of the candidates were selected based on their SED suggesting a cavity $<$10 au, the other four from a suggested inner cavity from low-resolution ALMA images. The continuum emission is analyzed through visibility analysis, using a power-law profile. Based on this work, we draw the following conclusions.

\begin{enumerate}
\item Five out of six SED-selected transition disk candidates with suggested small inner cavities $<$10 au do not show an indication of an inner dust cavity in their millimeter emission. 
\item Three of the SED-selected disks have a high inclination $>$65$^{\circ}$, which is demonstrated to cause a dip in the infrared part of the SED as well. However, high inclination also limits the detection of small cavities in the millimeter emission, also in the visibility plane.
\item Two of the SED-selected disks, Sz104 and Sz112, do not show an inner cavity in the millimeter distribution whereas the disks have modest inclinations, indicating that the dip in the SED may be the result of enhanced grain growth.
\item An inner dust cavity of 4 au was found for Sz~76, which makes this the only known transition disk in Lupus with a cavity that is consistent with classical photoevaporative clearing models. This makes this target the ideal testbed for photoevaporation models. The majority of the other resolved transition disks in Lupus can be explained with photoevaporation models under the assumption of carbon depletion and/or the inclusion of a dead zone. 
\item Considering our findings regarding the high inclination of SED-selected transition disks, dispersal timescales from transition disk fractions can only be derived from confirmed transition disk cavities resolved with millimeter imaging. 
\item Three of the four transition disk candidates selected from low-resolution imaging indeed contain a large inner cavity $\gtrsim$10 au, which brings the total number of transition disks with large inner cavities in Lupus to 13.
\item The discovery of two extremely compact disks (Sz~104 and Sz~112, with radii of 2.5 and 3.5 au, respectively) strengthens the possibility that at least some disks could be truly drift-dominated and may not contain substructure in the form of pressure bumps that limit the radial drift. The size-luminosity and $M_{\rm dust}-M_{\rm star}$ relations can be understood if the bright disks are dominated by disks with substructure, as previously inferred from multi-wavelength analysis by \citet{Tazzari2020}, whereas faint disks are dominated by drift-dominated disks.

\end{enumerate}

\begin{acknowledgements}
We would like to thank Barbara Ercolano for useful discussions  and Marco Tazzari for early help on the analysis.  
J.P.W. acknowledges support from NSF grant AST-1907486.
This project has received funding from the European Union's Horizon 2020 research and innovation programme under the Marie Sklodowska-Curie grant agreement No 823823 (DUSTBUSTERS).
This work was partly funded by the Deutsche Forschungsgemeinschaft (DFG, German Research Foundation) - 325594231.
ALMA is a partnership of ESO (representing its member states), NSF (USA) and NINS (Japan), together with NRC (Canada) and NSC and ASIAA (Taiwan) and KASI (Republic of Korea), in cooper- ation with the Republic of Chile. The Joint ALMA Observatory is operated by ESO, AUI/ NRAO and NAOJ. This paper makes use of the following ALMA data: 2018.1.01054.S, 2015.1.00222.S.
\end{acknowledgements}

\bibliographystyle{aa}

\newpage
\appendix

\section{Corner plots MCMC}
\label{sec:corners}
In this section we present the posteriors of the MCMC fitting of the intensity profiles of the disks in the sample. The best-fit results are listed in Table \ref{tbl:mcmc} and shown in Figures \ref{fig:viscurves}-\ref{fig:imagecomp}.

\begin{figure*}[!ht]
\includegraphics[width=\textwidth]{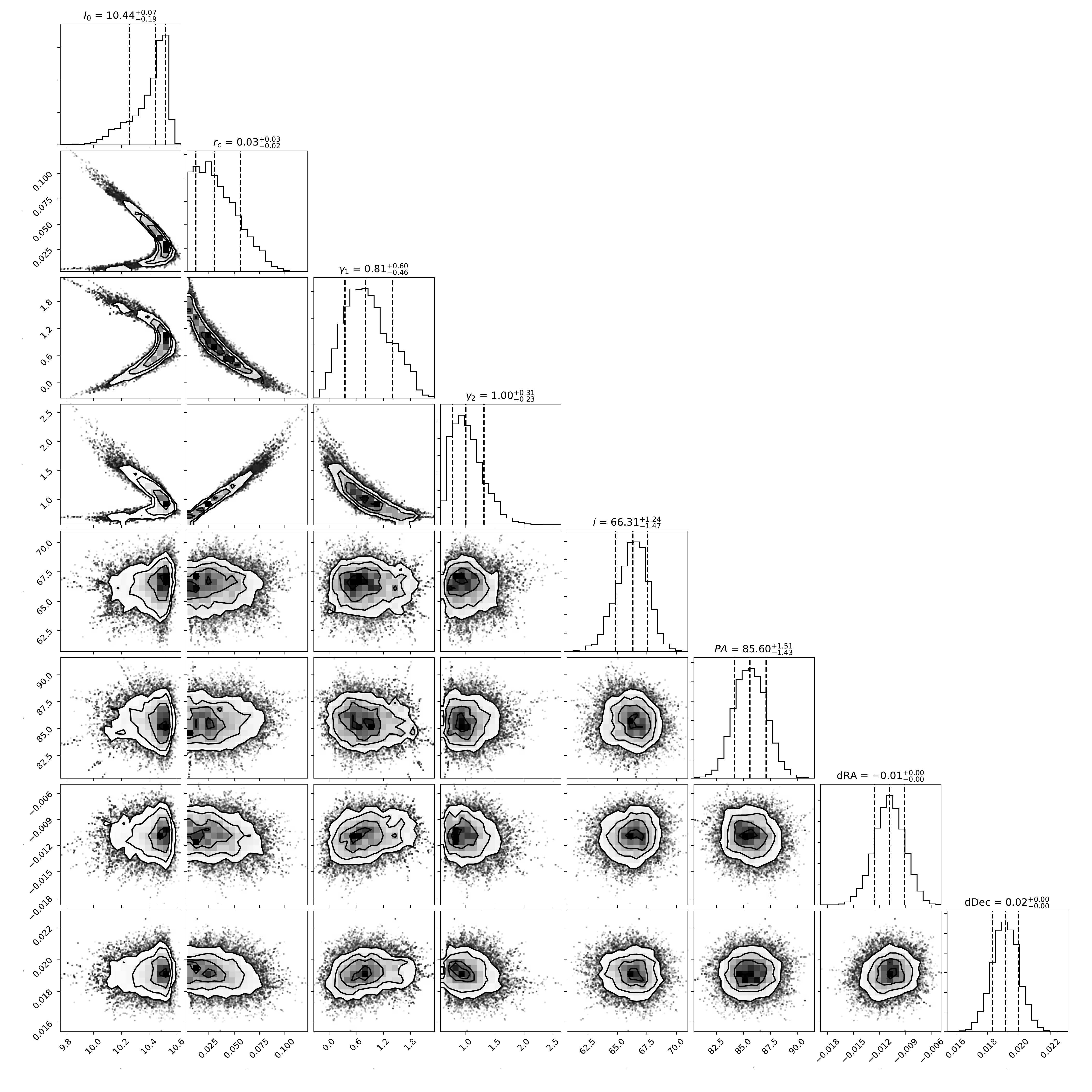}
\caption{Posteriors Sz76.}
\end{figure*}

\begin{figure*}[!ht]
\includegraphics[width=\textwidth]{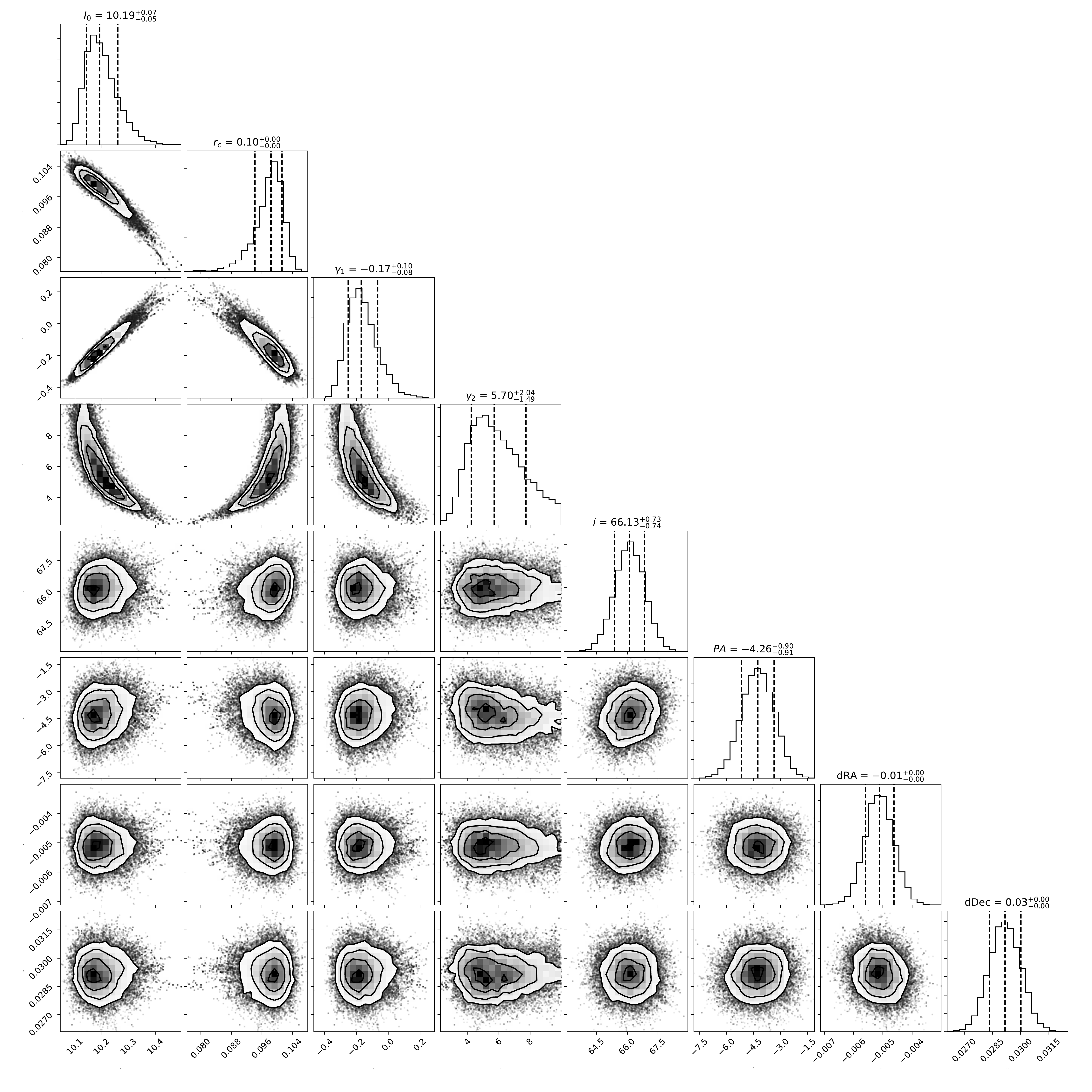}
\caption{Posteriors Sz103.}
\end{figure*}

\begin{figure*}[!ht]
\includegraphics[width=\textwidth]{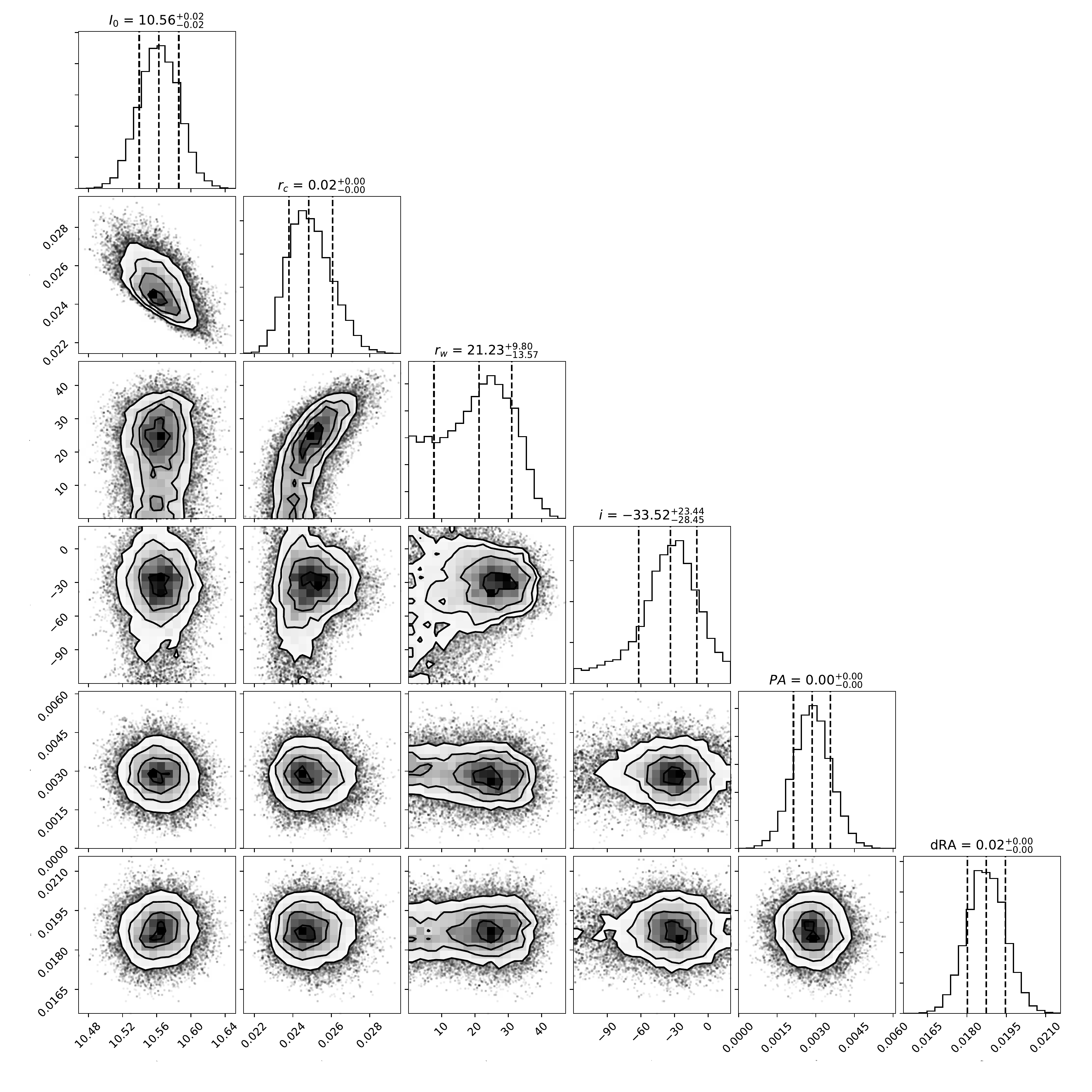}
\caption{Posteriors Sz112.}
\end{figure*}

\begin{figure*}[!ht]
\includegraphics[width=\textwidth]{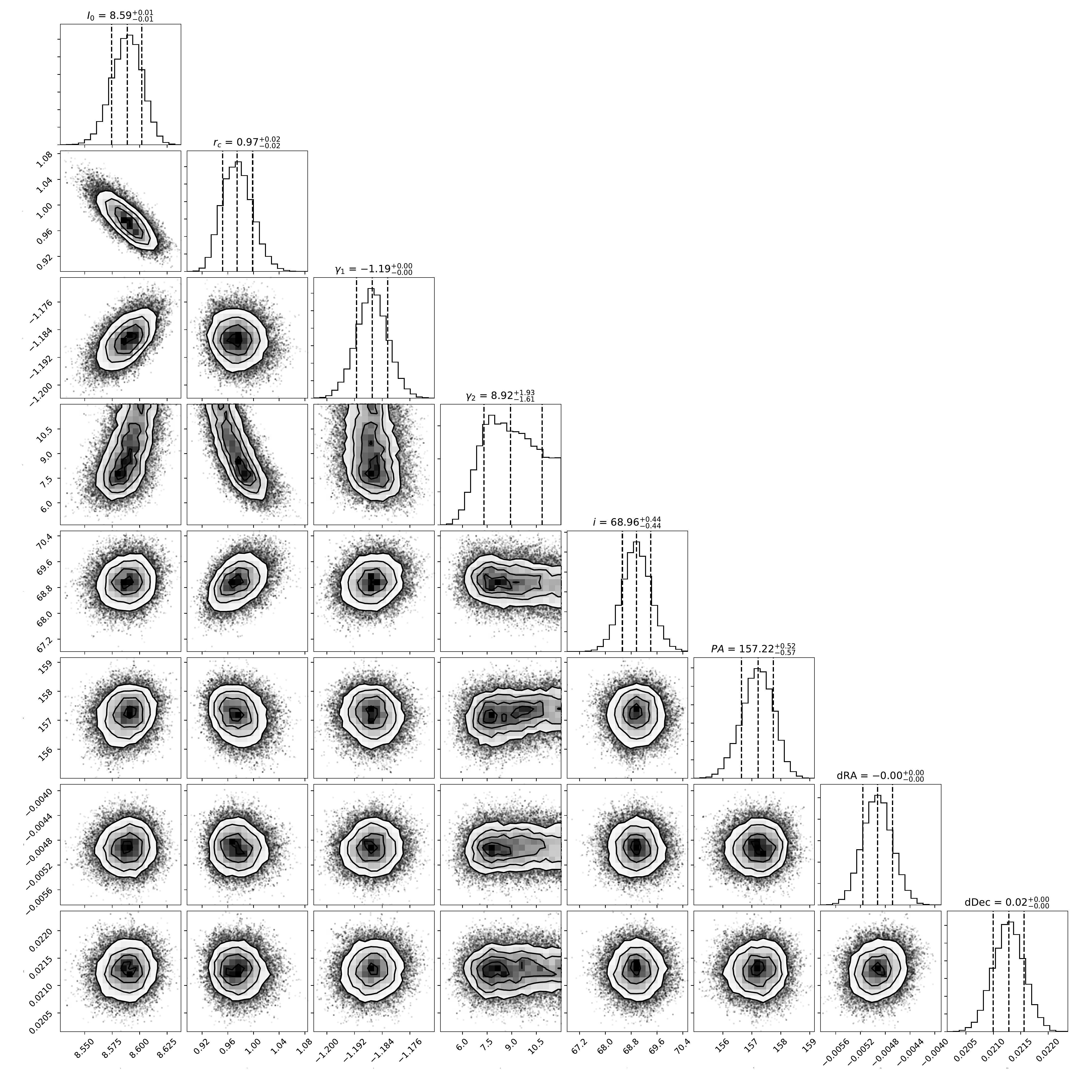}
\caption{Posteriors J16011549.}
\end{figure*}

\begin{figure*}[!ht]
\includegraphics[width=\textwidth]{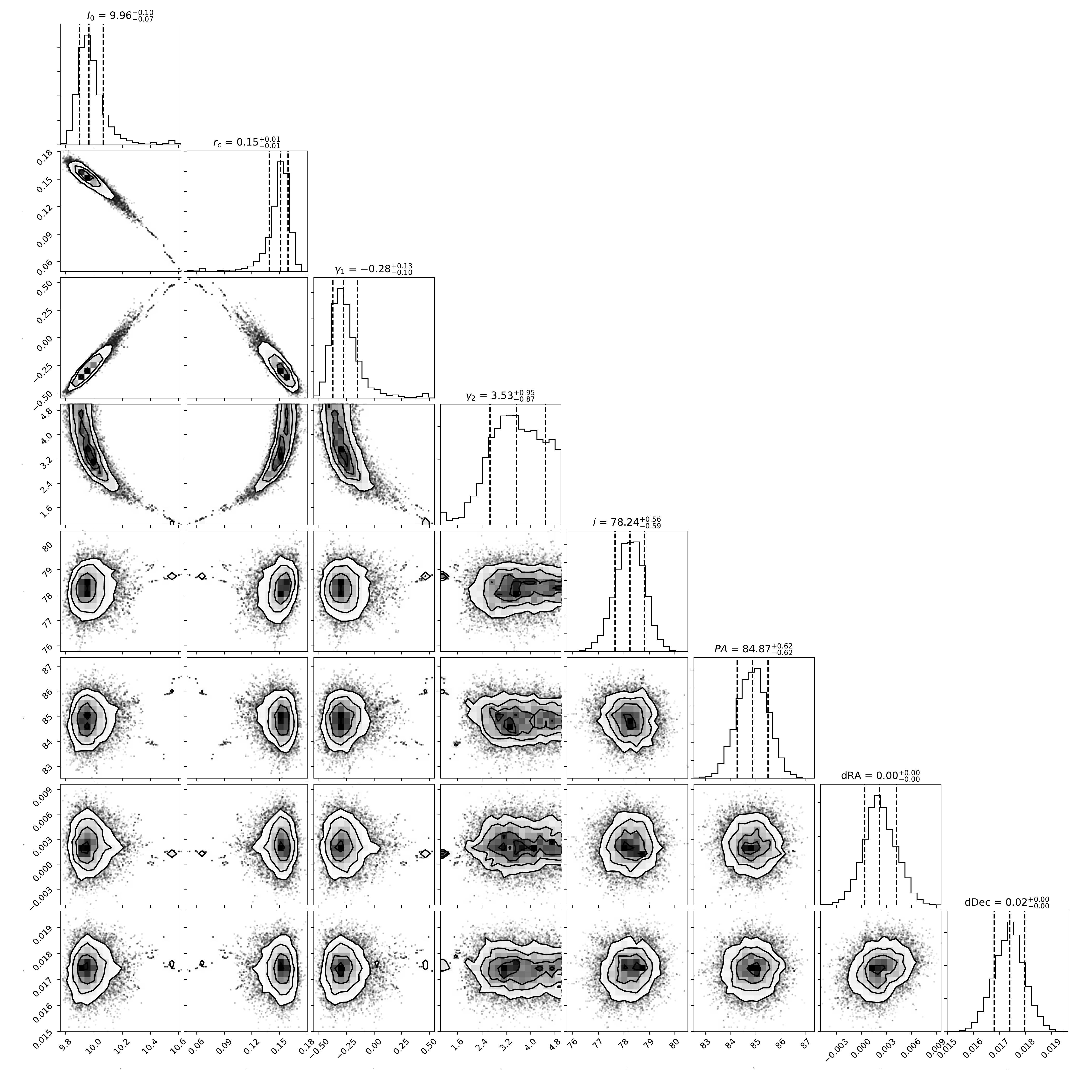}
\caption{Posteriors J16081497.}
\end{figure*}

\begin{figure*}[!ht]
\includegraphics[width=\textwidth]{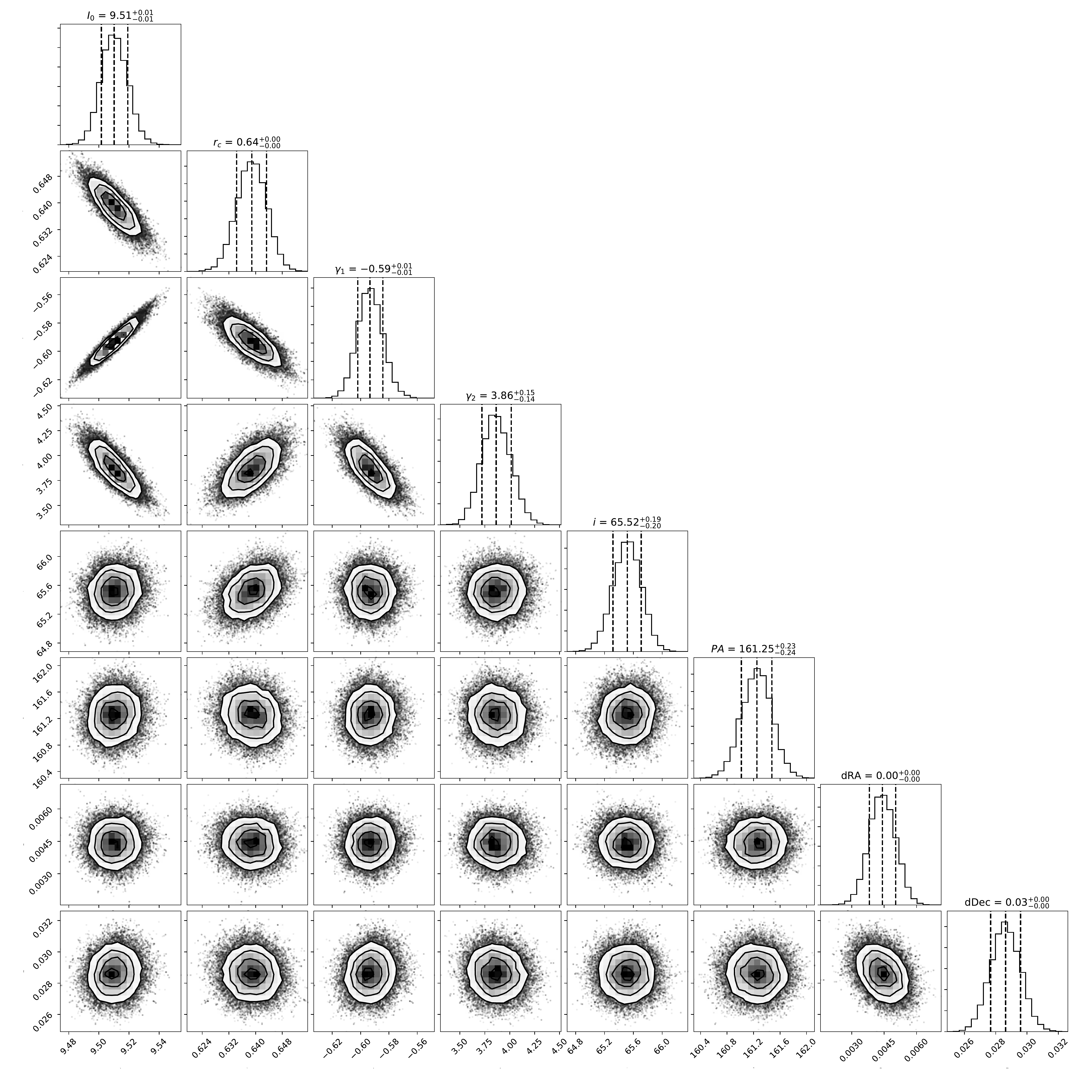}
\caption{Posteriors J16000236.}
\end{figure*}

\begin{figure*}[!ht]
\includegraphics[width=\textwidth]{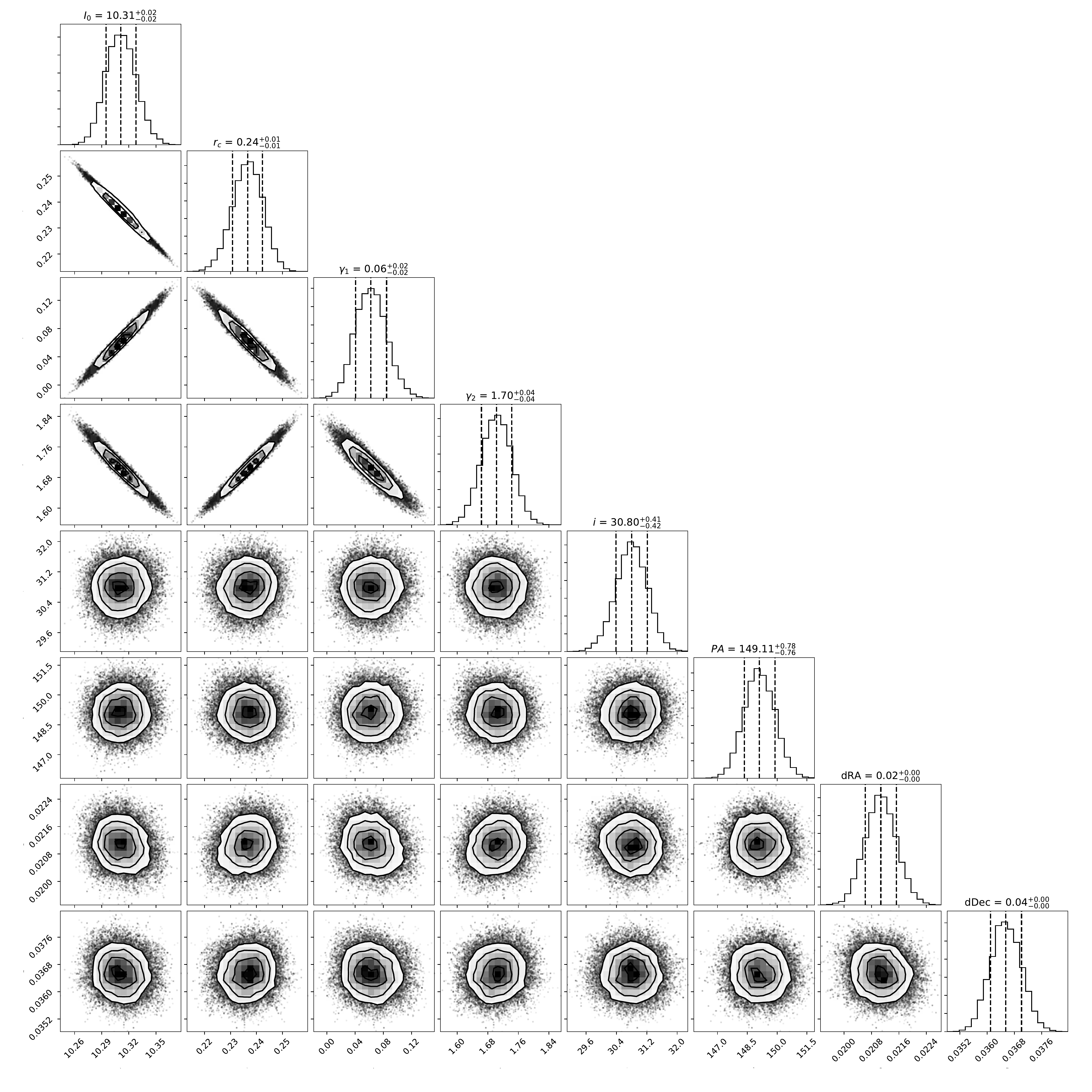}
\caption{Posteriors Sz129.}
\end{figure*}

\begin{figure*}[!ht]
\includegraphics[width=\textwidth]{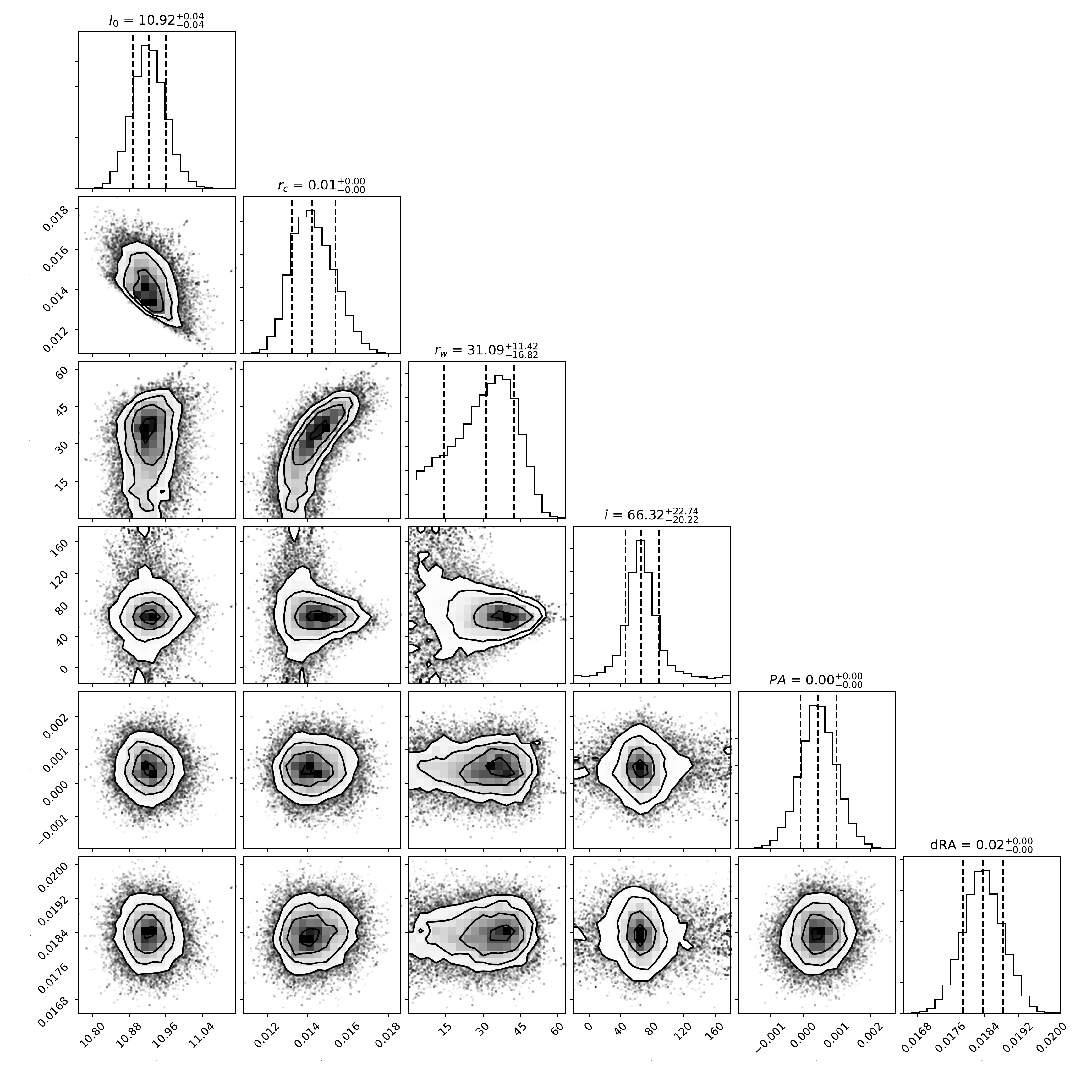}
\caption{Posteriors Sz104.}
\end{figure*}

\begin{figure*}[!ht]
\includegraphics[width=\textwidth]{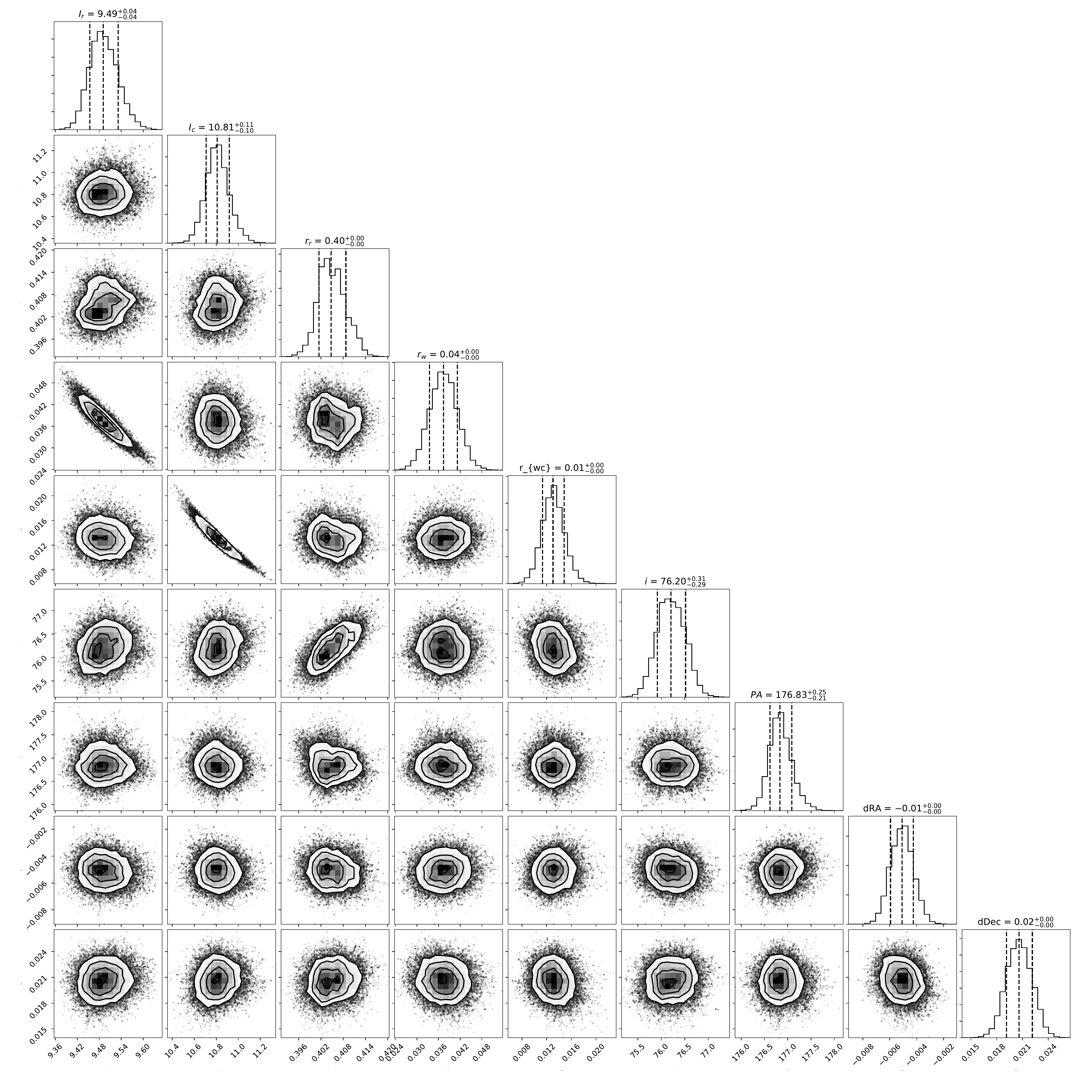}
\caption{Posteriors J16090141.}
\end{figure*}

\section{Large cavity transition disks}
\label{sec:largecav}
In this section the Band 6 intensity profiles of the large cavity transition disks from \citet{vanderMarel2018a} are fit using Galario, assuming an asymmetric radial Gaussian profile, following \citet{Pinilla2018tds} as these disks were not fit with this type of profile before:
\begin{align}
I(r) &= I_0 \exp\left(-\left(\frac{(r-r_{ring})^2}{2r_{w}^2}\right)\right)  \\
\textrm{with } r_w&=r_{wa} \textrm{ for } r<r_{ring} \\
 &=r_{wb} \textrm{ for } r>r_{ring}
\end{align}

\begin{figure*}[!ht]
\includegraphics[width=\textwidth,trim = 120 0 120 0]{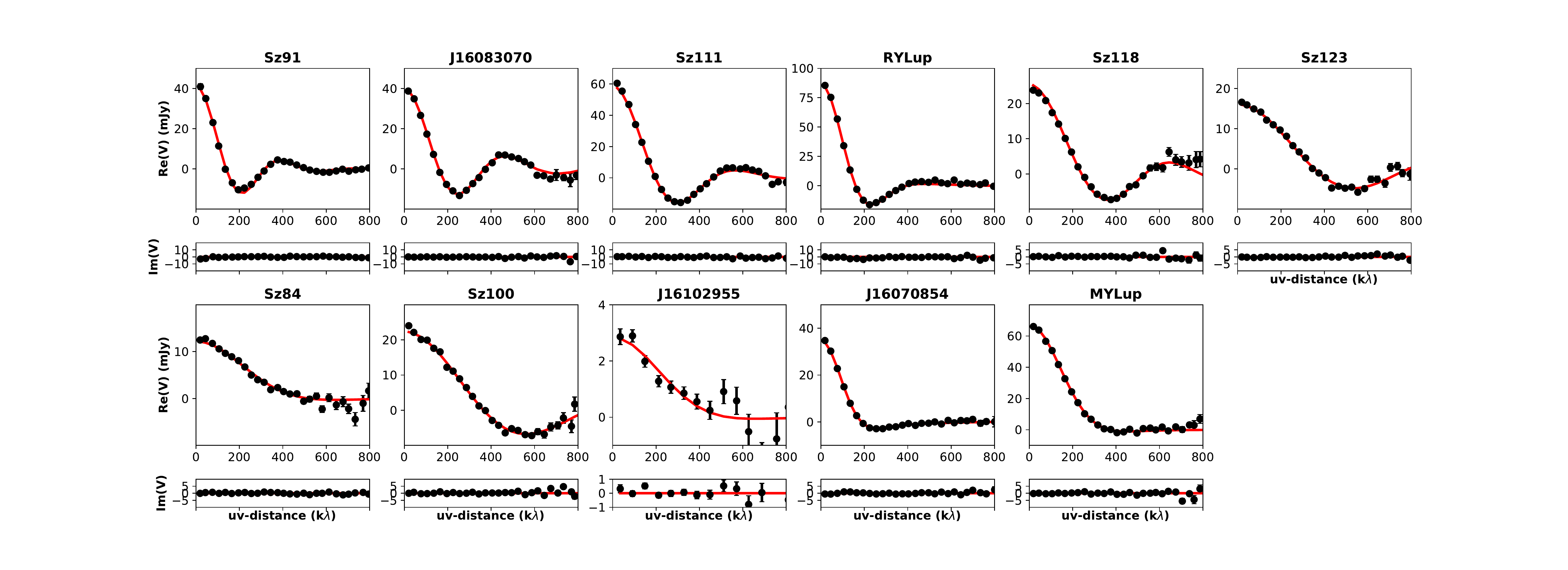}
\caption{Visibility curves and best-fit models of the large cavity transition disks from \citet{vanderMarel2018a}.}
\label{fig:largetd}
\end{figure*}

Together with the position angle, inclination and offsets in RA and Dec this implies a fit with 8 free parameters. Fitting is performed using MCMC with 80 walkers and 2000 steps, where the last 1000 steps are used to create the posteriors. For Sz~91, we use the Band 7 data from ALMA program 2012.1.00761.S presented in \citet{Tsukagoshi2019,Francis2020}, using the same fitting procedure. Table \ref{tbl:largetd} lists the best-fit parameters and uncertainties as derived from the posteriors, plus the dust ring radius in au and total flux from the shortest covered baselines. The visibility curves are shown in Figure \ref{fig:largetd}. This type of fitting had not been performed before for the full sample of transition disks in Lupus and is included here for comparison with the small cavity sample. We note that for Sz84, J16102955 and MYLup the presence of the cavity cannot be confirmed considering the high possible values of $r_{w1}$ compared to $r_c$, but with the large errorbar the cavity cannot be excluded either. High-resolution observations are required to confirm the presence of a cavity in these disks.

\begin{table*}[!ht]
\centering
\caption{Fit results of large cavity transition disks}
\label{tbl:largetd}
\footnotesize
\begin{tabular}{l|llllllll|lll}
\hline
Target & $\log I_0$ & $r_r$ & $r_{w1}$ & $r_{w2}$ & $i$ & $PA$ & dRA & dDec & $R_{\rm pk}$ & $R_{68}$ & $F_{1.3mm}$\\
& (Jy sr$^{-1}$)&(")&(")&(")&($^{\circ}$)&($^{\circ}$)&(")&(") & (au)&(au)&(mJy) \\
\hline
Sz91 &  9.39$^{+0.013}_{-0.012}$	&0.56$^{+0.014}_{-0.013}$	&0.12$^{+0.013}_{-0.013}$	&0.12$^{+0.010}_{-0.010}$	&47.66$^{+0.75}_{-0.76}$	&176.13$^{+1.02}_{-0.94}$	&-0.014$^{+0.004}_{-0.004}$&-0.001$^{+0.003}_{-0.003}$&	87 & 96	& 41\footnote{This is the 890 $\mu$m flux.} \\
J160830 & 10.02$^{+0.024}_{-0.023}$	&0.40$^{+0.017}_{-0.009}$ &0.013$^{+0.014}_{-0.008}$	&0.13$^{+0.014}_{-0.008}$	&72.86$^{+0.18}_{-0.17}$	&107.68$^{+0.18}_{-0.16}$	&-0.011$^{+0.002}_{-0.002}$&0.02$^{+0.001}_{-0.001}$&	62 & 81	& 40 \\
Sz111 & 9.90$^{+0.013}_{-0.008}$	&0.35$^{+0.010}_{-0.011}$ &0.064$^{+0.008}_{-0.011}$	&0.11$^{+0.007}_{-0.006}$	&54.38$^{+0.31}_{-0.46}$	&43.64$^{+0.40}_{-2.48}$	&-0.006$^{+0.001}_{-0.001}$&0.006$^{+0.001}_{-0.001}$&	55 & 67	& 58 \\
RYLup & 9.90$^{+0.029}_{-0.005}$	&0.42$^{+0.035}_{-0.009}$	&0.12$^{+0.008}_{-0.076}$	&0.20$^{+0.005}_{-0.116}$	&67.31$^{+0.36}_{-0.15}$	&108.91$^{+0.11}_{-0.30}$	&-0.001$^{+0.002}_{-0.002}$&0.06$^{+0.001}_{-0.009}$&	67 & 88	& 87 \\
Sz118 &  9.88$^{+0.035}_{-0.021}$	&0.39$^{+0.013}_{-0.034}$	&0.12$^{+0.013}_{-0.025}$	&0.015$^{+0.025}_{-0.010}$	&65.95$^{+0.62}_{-0.44}$	&172.85$^{+0.39}_{-0.41}$	&-0.007$^{+0.002}_{-0.002}$&0.04$^{+0.002}_{-0.002}$&	64 & 58	& 24 \\
Sz123A & 9.84$^{+0.051}_{-0.043}$	&0.26$^{+0.016}_{-0.027}$ &0.075$^{+0.014}_{-0.017}$	&0.019$^{+0.018}_{-0.011}$	&51.79$^{+0.91}_{-1.84}$	&154.45$^{+1.22}_{-5.21}$	&-0.040$^{+0.002}_{-0.002}$&0.034$^{+0.002}_{-0.003}$&	39 & 41	& 58 \\
Sz84 & 9.83$^{+0.050}_{-0.037}$	&0.081$^{+0.035}_{-0.031}$ &0.78$^{+1.52}_{-0.75}$	&0.13$^{+0.016}_{-0.020}$	&68.56$^{+1.25}_{-2.02}$	&169.3$^{+1.05}_{-1.32}$	&-0.033$^{+0.002}_{-0.002}$&-0.032$^{+0.003}_{-0.003}$&	12 & 26	& 12 \\
Sz100 &  9.97$^{+0.050}_{-0.038}$	&0.19$^{+0.031}_{-0.021}$ &0.029$^{+0.023}_{-0.018}$	&0.054$^{+0.01}_{-0.02}$	&42.89$^{+1.74}_{-1.10}$	&66.60$^{+1.51}_{-5.21}$	&-0.024$^{+0.002}_{-0.002}$&0.102$^{+0.002}_{-0.002}$&	26 & 31	& 22 \\
J161029 & 9.03$^{+0.64}_{-0.089}$	&0.07$^{+0.13}_{-0.053}$ &0.79$^{+1.48}_{-0.75}$	&0.13$^{+0.033}_{-0.090}$	&56.7$^{+25.3}_{-10.6}$	&106.8$^{+4.9}_{-9.6}$	&-0.070$^{+0.02}_{-0.02}$&0.002$^{+0.011}_{-0.011}$&	11 & 27	& 2.7 \\
J160708 & 9.62$^{+0.014}_{-0.012}$	&0.25$^{+0.026}_{-0.023}$ &0.093$^{+0.028}_{-0.025}$	&0.29$^{+0.01}_{-0.01}$	&73.08$^{+0.38}_{-0.29}$	&153.99$^{+0.36}_{-0.26}$	&-0.040$^{+0.002}_{-0.002}$&0.016$^{+0.003}_{-0.003}$&	40 & 77	& 35 \\
MYLup & 10.35$^{+0.009}_{-0.008}$	&0.10$^{+0.009}_{-0.010}$ &1.12$^{+1.26}_{-0.94}$	&0.20$^{+0.005}_{-0.004}$	&73.03$^{+0.19}_{-0.16}$	&59.13$^{+0.16}_{-0.14}$	&-0.020$^{+0.001}_{-0.001}$&0.025$^{+0.001}_{-0.001}$&	16 & 40	& 68 \\
\hline
\end{tabular}
\end{table*}




\end{document}